\documentclass[11pt]{article}
\usepackage{latexsym}
\usepackage{amssymb}
\usepackage{amsmath}
\usepackage{graphicx}
\usepackage{diagrams}
\usepackage{supertabular}
\usepackage{feynmp}
\usepackage{slash}

\def\CCC{{\mathbb C}}

\def\HHH{{\mathbb H}}

\def\QQQ{{\mathbb Q}} 
\def\RRR{{\mathbb R}}

\def\ZZZ{{\mathbb Z}} 

\def\C{{\cal C}} 
\def\D{{\cal D}} 
 
\def\F{{\cal F}} 
\def\G{{\cal G}}

\def\L{{\cal L}}

\def\O{{\cal O}}

\def\V{{\cal V}} 
\def\W{{\cal W}}

\def\Z{{\cal Z}} 

\def\({\left(}
\def\){\right)}
\def\[{\left[}
\def\]{\right]}
\def\<{\left\langle}
\def\>{\right\rangle}

\def\tensor{\otimes}

\def\d{\partial}

\def\lieO{{\rm O}}
\def\lieU{{\rm U}}

\def\lieSU{{\rm SU}}

\def\Aut{\mathop{\rm Aut}\nolimits}

\def\sign{\mathop{\rm sign}\nolimits}
\def\id{\mathop{\rm id}\nolimits}

\def\dist{{\sf Dist}}
\def\feyn{{\sf Feyn}}
\def\lag{{\sf Lag}}
\def\wf{{\sf WF}}

\def\Clif{\mathop{\rm C}\nolimits}
\def\mx{\mathop{\rm M}\nolimits}

\newenvironment{centre}{\begin{center}}{\end{center}}
\newenvironment{itemise}{\begin{itemize}}{\end{itemize}}

\begin{document}

\begin{fmffile}{qft_pics}

\begin{centre}
\huge {\bf QUANTUM FIELD THEORY} \\
\Large Notes taken from a course of R. E. Borcherds, Fall 2001, Berkeley\\
\end{centre}

Richard E. Borcherds,\\
Mathematics department, Evans Hall, UC Berkeley, CA 94720, U.S.A.\\
e--mail: {\tt reb@math.berkeley.edu}\\
home page: {\tt www.math.berkeley.edu/\~{}reb}

Alex Barnard,\\
Mathematics department, Evans Hall, UC Berkeley, CA 94720, U.S.A.\\
e--mail: {\tt barnard@math.berkeley.edu}\\
home page: {\tt www.math.berkeley.edu/\~{}barnard}

\vskip 18mm
\tableofcontents

\eject\section{Introduction}
\subsection{Life Cycle of a Theoretical Physicist}

\begin{enumerate}
\item Write down a \emph{Lagrangian density} $L$.  This is a polynomial in
\emph{fields} $\psi$ and their derivatives.  For example
$$L[\psi] = \d_\mu\psi\d^\mu\psi - m^2\psi^2 + \lambda\psi^4$$
\item Write down the \emph{Feynman path integral}.  Roughly speaking
this is
$$\int e^{i\int L[\psi]} \D\psi$$
The value of this integral can be used to compute ``cross sections''
for various processes.
\item Calculate the Feynman path integral by expanding as a formal
power series in the ``coupling constant'' $\lambda$.
$$a_0 + a_1\lambda + a_2\lambda + \cdots$$
The $a_i$ are finite sums over \emph{Feynman diagrams}.  Feynman
diagrams are a graphical shorthand for finite dimensional integrals.
\item Work out the integrals and add everything up.
\item Realise that the finite dimensional integrals do not converge.
\item Regularise the integrals by introducing a ``cutoff'' $\epsilon$
(there is usually an infinite dimensional space of possible
regularisations).  For example
$$\int_\RRR {1\over x^2} dx \longrightarrow \int_{|x|>\epsilon}
{1\over x^2} dx$$
\item Now we have the series
$$a_0(\epsilon) + a_1(\epsilon)\lambda + \cdots$$
{\bf Amazing Idea:} Make $\lambda$, $m$ and other parameters of the
Lagrangian depend on $\epsilon$ in such a way that terms of the
series are independent of $\epsilon$.
\item Realise that the new sum still diverges even though we have made
all the individual $a_i$'s finite.  \emph{No good way of fixing this
is known}.  It appears that the resulting series is in some sense
an asymptotic expansion.
\item Ignore step 8, take only the first few terms and compare with
experiment.
\item Depending on the results to step 9: Collect a Nobel prize or
return to step 1.
\end{enumerate}

There are many problems that arise in the above steps

\begin{description}
\item[Problem 1] The Feynman integral is an integral over an infinite
dimensional space and there is {\bf no} analogue of Lebesgue measure.
\item[Solution] Take what the physicists do to evaluate the integral
as its definition.
\item[Problem 2] There are many possible cutoffs.  This means the
value of the integral depends not only on the Lagrangian but also on
the choice of cutoff.
\item[Solution] There is a group $G$ called the \emph{group of finite
renormalisations} which acts on both Lagrangians and cutoffs.  QFT is
unchanged by the action of $G$ and $G$ acts transitively on the space
of cutoffs.  So, we only have to worry about the space of Lagrangians.
\item[Problem 3] The resulting formal power series (even after
renormalisation) does not converge.
\item[Solution] Work in a formal power series ring.
\end{description}

\subsection{Historical Survey of the Standard Model}

\begin{itemise}
\item[1897] Thompson discovered the electron.  This is the first
elementary particle to be discovered.

Why do we believe the electron is elementary?  If it were composite
then it would be not point-like and it could also vibrate.  However
particle experiments have looked in detail at electrons and they still
seem point-like.  Electrons have also been bashed together extremely
hard and no vibrations have ever been seen.
\item[1905] Einstein discovers photons.  He also invents special
relativity which changes our ideas about space and time.
\item[1911] The first scattering experiment was performed by
Rutherford; he discovered the nucleus of the atom.
\item[1925] Quantum mechanics is invented and physics becomes
impossible to understand.
\item[c1927] Quantum field theory is invented by Jordan, Dirac, ...
\item[1928] Dirac invents the wave equation for the electron and
predicts positrons.  This is the first particle to be predicted from a
theory.
\item[1929] Heisenberg and Pauli notice QFT has lots of infinities.
The types that occur in integrals are called ``ultraviolet'' and
``infrared''.

UV divergences occur when local singularities are too
large to integrate.  For example the following integral has a UV
divergence at the origin for $s\ge 1$
$$\int_\RRR {1 \over x^s}\ dx$$
IR divergences occur when the behaviour at
$\infty$ is too large.  For example the above integral has an IR
divergence for $s\le 1$.

Finally even when these are removed the resulting power series doesn't
converge.
\item[1930] Pauli predicts the neutrino because the decay
$$n \longrightarrow p + e^-$$
has a continuous spectrum for the electron.
\item[1932] Chadwick detects the neutron.

Positrons are observed.
\item[1934] Fermi comes up with a theory for $\beta$-decay.
$\beta$-decay is involved in the following process
$$n \longrightarrow p + e^- + \overline{\nu}_e$$
Yukawa explains the ``strong'' force in terms of three mesons $\pi^+$,
$\pi^0$ and $\pi^-$.  The idea is that nucleons are held together by
exchanging these mesons.  The known range of the strong force allows
the masses of the mesons to be predicted, it is roughly 100MeV.
\item[1937] Mesons of mass about 100MeV are detected in cosmic rays.
The only problem is that they don't interact with nucleons!
\item[1947] People realised that the mesons detected were not Yukawa's
particles (now called pions).  What was detected were muons (these are
just like electrons only heavier).
\item[1948] Feynman, Schwinger and Tomonaga independently invent a
systematic way to control the infinities in QED.  Dyson shows that
they are all equivalent.  This gives a workable theory of QED.
\item[1952] Scattering experiments between pions and nucleons find
resonances.  This is interpreted as indicating the existence of a new
particle
$$n+\pi \longrightarrow \hbox{new particle} \longrightarrow n+\pi$$
A huge number of particles were discovered like this throughout the
50's.
\item[1956] Neutrinos are detected using very high luminosity beams
from nuclear reactors.

Yang-Mills theory is invented.
\item[1957] Alvarez sees cold fusion occurring in bubble chambers.
This reaction occurs because a proton and a muon can form a ``heavy
hydrogen atom'' and due to the decrease in orbital radius these can
come close enough to fuse.

Parity non-conservation is noticed by Lee-Yang and Wu.  Weak
interactions are not symmetric under reflection.  The experiment uses
cold ${}^{60}\hbox{Co}$ atoms in a magnetic field.  The reaction
$${}^{60}\hbox{Co} \longrightarrow {}^{60}\hbox{Ni} + \nu_e + e^-$$
occurs but the electrons are emitted in a preferred direction.
\item[1960s] Gell-Mann and Zweig predict quarks.  This explains the
vast number of particles discovered in the 1950's.  There are 3 quarks
(up, down and strange).  The observed particles are not elementary but
made up of quarks.  They are either quark-antiquark pairs (denoted by
$q\bar q$) or quark triplets (denoted by $qqq$).  This predicts 
a new particle (the $\Omega^-$ whose quark constituents are $sss$).
\item[1968] Weinberg, Salem and Glashow come up with the electroweak
theory.  Like Yukawa's theory there are three particles that propagate
the weak force.  These are called ``intermediate vector bosons''
$W^+$, $W^-$ and $Z^0$.  They have predicted masses in the region of
80GeV.
\item[1970] Iliopoulos and others invent the charm quark.
\item[1972] 't Hooft proves the renormalisability of gauge theory.

Kobayaski and Maskawa predict a third generation of elementary
particles to account for CP violation.  The first generation consists
of $\{u,d,e,\nu_e\}$, the second of $\{c,s,\mu,\nu_\mu\}$ and the
third of $\{t,b,\tau,\nu_\tau\}$.
\item[1973] Quantum chromodynamics is invented.  ``Colour'' is needed
because it should not be possible to have particles like the
$\Omega^-$ (which is made up of the quark triplet $sss$) due to the
Pauli exclusion principle.  So quarks are coloured in 3 different
colours (usually red, green and blue).  QCD is a gauge theory on
$\lieSU(3)$.  \emph{The three dimensional vector space in the gauge
symmetry is the colour space}.
\item[1974] The charm quark is discovered simultaneously by two
groups.  Ting fired protons at a Beryllium target and noticed a sharp
resonance in the reaction at about 3.1GeV.  SPEAR collided electrons
and positrons and noticed a resonance at 3.1GeV.

Isolated quarks have never been seen.  This is called ``quark
confinement''.  Roughly speaking, the strong force decays very slowly
with distance so it requires much more energy to separate two quarks
by a large distance than it does to create new quark pairs and
triples.  This is called the ``asymptotic freedom'' of the strong
force.
\item[1975] ``Jets'' are observed.  Collision experiments often emit
the resulting debris of particles in narrow beams called jets.  This
is interpreted as an observation of quarks.  The two quarks start
heading off in different directions and use the energy to repeatedly
create new quark pairs and triples which are observed as particles.

The $\tau$ particle is detected.
\item[1977] Ledermann discovers the upsilon ($\Upsilon$) particle
which consists of the quarks $b\bar b$.
\item[1979] 3-jet events are noticed.  This is seen as evidence for
the gluon.  Gluons are the particles which propagate the inter-quark force.
\item[1983] CERN discover the intermediate vector bosons.
\item[1990] It was argued that because neutrinos were massless there
could only be 3 generations of elementary particles.  This no longer
seems to be valid as neutrinos look like they have mass.
\end{itemise}
\eject
\subsection{Some Problems with Neutrinos}

Neutrinos can come from many different sources
\begin{itemise}\parskip 0pt
\item Nuclear reactors
\item Particle beams
\item Leftovers from the big bang
\item Cosmic rays
\item Solar neutrinos
\end{itemise}
For cosmic rays there processes that occur are roughly
\begin{eqnarray*}
\alpha + \hbox{atom} &\longrightarrow& \pi^- + \cdots\\
\pi^- &\longrightarrow& \mu^- + \overline{\nu}_\mu \\
\mu^- &\longrightarrow& e^- + \overline{\nu}_e + \nu_\mu
\end{eqnarray*}
This predicts that there should be equal numbers of electron, muon and
anti-muon neutrinos.  However there are too few muon neutrinos by a
factor of about 2.  Also there seems to be a directional dependence to
the number of $\nu_\mu$'s.  There are less $\nu_\mu$'s arriving at a
detector if the cosmic ray hit the other side of the earth's
atmosphere.  This suggests that the numbers of each type of neutrino
depend on the time they have been around.

Inside the sun the main reactions that take place are
\begin{eqnarray*}
p+\bar p &\longrightarrow& {}^2\hbox{H} + e^+ + \nu_e \\
p+e^-+p &\longrightarrow& {}^2\hbox{H} + \nu_e \\
{}^2\hbox{H} + p &\longrightarrow& {}^3\hbox{He} + \gamma \\
{}^3\hbox{He} + {}^3\hbox{He} &\longrightarrow& {}^4\hbox{He} + p
+ p \\
{}^3\hbox{He} + {}^4\hbox{He} &\longrightarrow& {}^7\hbox{Be} + p \\
{}^7\hbox{Be} + e^- &\longrightarrow& {}^7\hbox{Be} + \nu_e \\
{}^7\hbox{Be} + e^- &\longrightarrow& \left\{ \begin{array}{l}
{}^7\hbox{Li} + \nu_e \\ {}^7\hbox{Li}^* + \nu_e\end{array}\right. \\
{}^7\hbox{Be} + p &\longrightarrow& {}^8\hbox{B} + \gamma \\
{}^8\hbox{B} &\longrightarrow& {}^8\hbox{Be} + e^+ + \nu_e \\
{}^7\hbox{Li} + p &\longrightarrow& {}^4\hbox{He} + {}^4\hbox{He}\\
{}^8\hbox{Be} &\longrightarrow& {}^4\hbox{He} + {}^4\hbox{He}
\end{eqnarray*}
All of these are well understood reactions and the spectra of
neutrinos can be predicted.  However comparing this with experiment
gives only about $1\over 3$ the expected neutrinos.
\footnote{
The amount of ${}^{12}\hbox{C}$ produced can also be predicted and it
is much smaller than what is seen in nature (e.g. we exist).  Hoyle's
amazing idea (amongst many crazy ones) was that there should be an
excited state of ${}^{12}\hbox{C}$ of energy 7.45MeV.  This was later
detected.}

That the number of neutrinos detected in solar radiation is about
$1\over 3$ of what is expected is thought to be an indication that
neutrinos can oscillate between different generations.

\subsection{Elementary Particles in the Standard Model}

Below is a list of the elementary particles that make up the standard
model with their masses in GeV.

{\bf FERMIONS}

These are the fundamental constituents of matter.  All have spin
$-{1\over 2}$.

\begin{centre}
{\it Quarks}
$$
\hbox{charge $2\over 3$ }
\left\{
\begin{array}{rlcrl}
up&0.003&\qquad&down&0.006 \\
charm&1.3& &strange&0.1 \\
top&175& &bottom&4.3
\end{array}
\right\}
\hbox{ charge $-{1\over3}$}
$$
\end{centre}
\begin{centre}
{\it Leptons}
$$
\hbox{charge $-1$ }
\left\{
\begin{array}{rlcrl}
e&0.0005&\qquad&\nu_e& <0.00000001 \\
\mu&0.1& &\nu_\mu& <0.0002 \\
\tau&1.8& &\nu_\tau& <0.02
\end{array}
\right\}
\hbox{ charge $0$}
$$
\end{centre}
{\bf BOSONS}

These are the particles that mediate the fundamental forces of
nature.  All have spin $0$.  Their charge (if any) is indicated in the
superscript.

\begin{centre}
\begin{minipage}[t]{1.5in}
\begin{centre}
{\it Electro--Magnetic}
$$
\begin{array}{ll}
\gamma&0
\end{array}
$$
\end{centre}
\end{minipage}
\begin{minipage}[t]{1.5in}
\begin{centre}
{\it Weak}
$$
\begin{array}{ll}
W^+&80 \\
W^-&80 \\
Z^0&91
\end{array}
$$
\end{centre}
\end{minipage}
\begin{minipage}[t]{1.5in}
\begin{centre}
{\it Strong}
$$
\begin{array}{ll}
g&0
\end{array}
$$
\end{centre}
\end{minipage}
\end{centre}

\eject\section{Lagrangians}
\subsection{What is a Lagrangian?}

For the abstract setup we have
\begin{itemise}\parskip 0pt
\item A finite dimensional vector space $\RRR^n$.  This is {\bf spacetime}.
\item A finite dimensional complex vector space $\Phi$ with a complex
conjugation $\star$.  This is the {\bf space of abstract fields}.
\end{itemise}

$\RRR^n$ has an $n$-dimensional space of translation invariant vector
fields denoted by $\d_\mu$.  These generate a polynomial ring of
differential operators $\RRR[\d_1,\dots,\d_n]$.  We can apply the
differential operators to abstract fields to get $\RRR[\d_1,\dots,
\d_n] \tensor \Phi$.  Finally we take
$$\V = \hbox{Sym}\[\RRR[\d_1,\dots,\d_n]\tensor \Phi\]$$
This is the space of Lagrangians.  Any element of it is called a
Lagrangian.

Note, we can make more general Lagrangians than this by taking the
formal power series rather than polynomial algebra.  We could also
include fermionic fields and potentials which vary over spacetime.

A {\bf representation of abstract fields} maps each element of $\Phi$ to a
complex function on spacetime (preserving complex conjugation).  Such
a map extends uniquely to a map from $\V$ to complex functions on
spacetime preserving products and differentiations.

Fields could also be represented by operators on a Hilbert space or
operator--valued distributions although we ignore this while
discussing classical field theory.

\subsection{Examples}
\begin{enumerate}
\item \emph{The Lagrangian of a vibrating string}

$\Phi$ is a one dimensional space with basis $\varphi$ such that
$\varphi^\star = \varphi$.  Spacetime is $\RRR^2$ with coordinates
$(x,t)$.  The Lagrangian is
$$\(\d\varphi \over \d t\)^2 - \(\d\varphi \over \d x\)^2$$
How is the physics described by the Lagrangian?  To do this we use
{\bf Hamilton's principle} which states that classical systems evolve
so as to make the action (the integral of the Lagrangian over
spacetime) stationary.

If we make a small change in $\varphi$ we will get an equation which
tells us the conditions for the action to be stationary.  These
equations are called the {\bf Euler--Lagrange equations}.
\begin{eqnarray*}
\delta\int L &=& \int\( 2{\d\varphi \over \d t} {\d\delta\varphi
\over \d t} - 2{\d\varphi \over \d x}{\d\delta\varphi \over \d x} \)
dx\ dt \\
&=& \int\( -2{\d^2\varphi \over \d t^2} + 2{\d^2\varphi \over \d x^2}
\)\delta\varphi\ dx\ dt
\end{eqnarray*}
Hence the Euler--Lagrange equations for this Lagrangian are
$${\d^2\varphi \over \d t^2} = {\d^2\varphi \over \d x^2}$$
This PDE is called the \emph{wave equation}.
\item \emph{The real scalar field}

$\Phi$ is a one dimensional space with basis $\varphi$ such that
$\varphi^\star = \varphi$.  Spacetime is $\RRR^n$.  The Lagrangian is
$$\d_\mu \varphi \d^\mu \varphi + m^2\varphi^2$$
Proceeding as before we derive the Euler--Lagrange equations
$$\d^\mu \d_\mu \varphi = m^2\varphi$$
This PDE is called the \emph{Klein--Gordon equation}.
\item \emph{A non-linear example: $\varphi^4$ theory}

The setup is exactly like the Klein--Gordon equation except that the
Lagrangian is
$$\d_\mu \varphi \d^\mu \varphi + m^2\varphi^2 + \lambda\varphi^4$$
This leads to the Euler--Lagrange equations
$$\d^\mu \d_\mu \varphi = m^2\varphi + 2\lambda\varphi^3$$
\item \emph{The complex scalar field}

This time $\Phi$ is two dimensional with basis $\varphi$ and
$\varphi^\star$.  Spacetime is $\RRR^n$ and the Lagrangian is
$$\d_\mu\varphi^\star \d^\mu \varphi + m^2\varphi^\star \varphi$$
The Euler--Lagrange equations are
$$\d^\mu \d_\mu \varphi^\star = m^2 \varphi^\star \atop
\d^\mu \d_\mu \varphi = m^2 \varphi$$
\end{enumerate}

\subsection{The General Euler--Lagrange Equation}

The method to work out the Euler--Lagrange equations can easily be
applied to a general Lagrangian which consists of fields and their
derivatives.  If we do this we get
$${\d L \over \d\varphi} - \d_\mu {\d L \over \d(\d_\mu \varphi)} +
\d_\mu \d_\nu {\d L \over \d(\d_\mu \d_\nu \varphi)} - \cdots = 0$$
We get $\dim(\Phi)$ of these equations in general.

The Euler--Lagrange equations and all their derivatives generate an
ideal $EL$ in $\V$ and (much as in the theory of $\D$--modules) we can
give a purely algebraic way to study solutions to the Euler--Lagrange
equations.  The quotient $\V/EL$ is a ring over $\CCC[\d_\mu]$.
Similarly the space of complex functions on spacetime $\RRR$,
$C^\infty(\RRR)$, is a ring over $\CCC[\d_\mu]$.  Ring homomorphisms
over $\CCC[\d_\mu]$ from $\V / EL$ to $C^\infty(\RRR)$ then correspond
one-to-one with solutions to the Euler--Lagrange equations.

\eject\section{Symmetries and Currents}

Consider the complex scalar field.  If we set
$$j^\mu = \varphi^\star \d^\mu \varphi - \varphi \d^\mu
\varphi^\star$$
then it is easy to check that
$$\d_\mu j^\mu = 0 \in \V/EL$$
It is also easy to check that the following transformations (if
performed simultaneously) preserve the Lagrangian
$$\varphi \longrightarrow e^{i\theta}\varphi \qquad\qquad\qquad
\varphi^\star \longrightarrow e^{-i\theta}\varphi^\star$$
The aim of this section is to explain how these two facts are
related and why $j^\mu$ is called a {\bf conserved current}.

\subsection{Obvious Symmetries}

Define $\Omega^1(\V)$ to be the module over $\V$ with basis
$\delta\varphi$, $\delta\d_\mu\varphi$, ...  There is a natural map
$$\delta : \V \longrightarrow \Omega^1(\V)$$
such that
$$\delta(ab) = (\delta a)b + a(\delta b)\qquad\hbox{and}\qquad
\delta\d_\mu = \d_\mu \delta$$
We can regard the operator $\delta$ as performing a general infinitesimal
deformation of the fields in the Lagrangian.  We now work out $\delta
L$ and use the Leibnitz rule to make sure that no elements $\delta \varphi$
are ever directly differentiated by $\d_\mu$ ({\it cf} integration by
parts).  We get an expression of the form
$$\delta L = \delta\varphi \cdot A + \delta\varphi^\star \cdot B +
\d_\mu J^\mu$$
where $A$ and $B$ are the Euler--Lagrange equations obtained by
varying the fields $\varphi$ and $\varphi^\star$.

Suppose that we have an infinitesimal symmetry which preserves $L$.
This means that if we replace $\delta\varphi$ and $\delta\varphi^\star$ in the
above expression by the infinitesimal symmetry generators we get
$\delta L=0$.  Then
$$- \delta\varphi \cdot A - \delta\varphi^\star \cdot B = \d_\mu j^\mu$$
where $j^\mu$ is what is obtained from $J^\mu$ by replacing $\delta\varphi$
and $\delta\varphi^\star$ by the infinitesimal symmetry generators.  In
$\V/EL$ this becomes
$$\d_\mu j^\mu = 0$$
because $A$ and $B$ are in $EL$.
To illustrate this, consider the complex scalar field with Lagrangian
$$L = \d_\mu\varphi^\star \d^\mu\varphi + m^2\varphi^\star \varphi$$
Hence
\begin{eqnarray*}
\delta L &=& \d_\mu \delta \varphi^\star\d^\mu\varphi + \d_\mu\varphi^\star
\d^\mu \delta \varphi + m^2\delta \varphi^\star \varphi + m^2\varphi^\star
\delta \varphi \\
&=& \d_\mu \(\delta \varphi^\star\d^\mu\varphi\) -
\delta \varphi^\star\d_\mu\d^\mu \varphi + \d^\mu\(
\d_\mu\varphi^\star \delta \varphi\) -
\d^\mu\d_\mu\varphi^\star \delta \varphi + \\
&& + m^2\delta \varphi^\star \varphi + m^2\varphi^\star \delta \varphi \\
&=& \delta \varphi^\star \cdot (m^2\varphi - \d_\mu\d^\mu\varphi) +
\delta \varphi \cdot (m^2\varphi^\star - \d_\mu\d^\mu\varphi^\star) +
\d_\mu \(\delta \varphi^\star\d^\mu\varphi + \d^\mu\varphi^\star
\delta \varphi\)
\end{eqnarray*}
We can see both Euler--Lagrange equations in the final expression for
$dL$.  The symmetry was
$$\varphi\longrightarrow e^{i\theta}\varphi \qquad\qquad\qquad
\varphi^\star\longrightarrow e^{-i\theta}\varphi^\star$$
whose infinitesimal generators are
$$\delta\varphi = i\varphi \qquad\qquad\qquad \delta\varphi^\star =
-i\varphi^\star$$
Substituting this into the final expression for $\delta L$ we easily check
that $\delta L=0$.  The conserved current
$$j^\mu = \varphi^\star \d^\mu \varphi - \varphi \d^\mu
\varphi^\star$$
is obtained from the $\d_\mu(*)$ term in $\delta L$ by substituting the
above infinitesimal symmetries (and dividing by $-i$).

Thus, given any infinitesimal symmetry of the Lagrangian $L$
we can associate something called $j^\mu$ which satisfies the above
differential equation.  $j^\mu$ is called a {\bf Noether current}.
$j^\mu$ is really a closed $(n-1)$--form given
by
$$\omega = j^1 dx^2\wedge dx^3\wedge\cdots - j^2 dx^1\wedge dx^3\wedge
\cdots + \cdots$$
The differential equation that $j^\mu$ satisfies becomes the
closedness condition
$$d\omega = 0$$

\begin{centre}
\includegraphics[scale=.5]{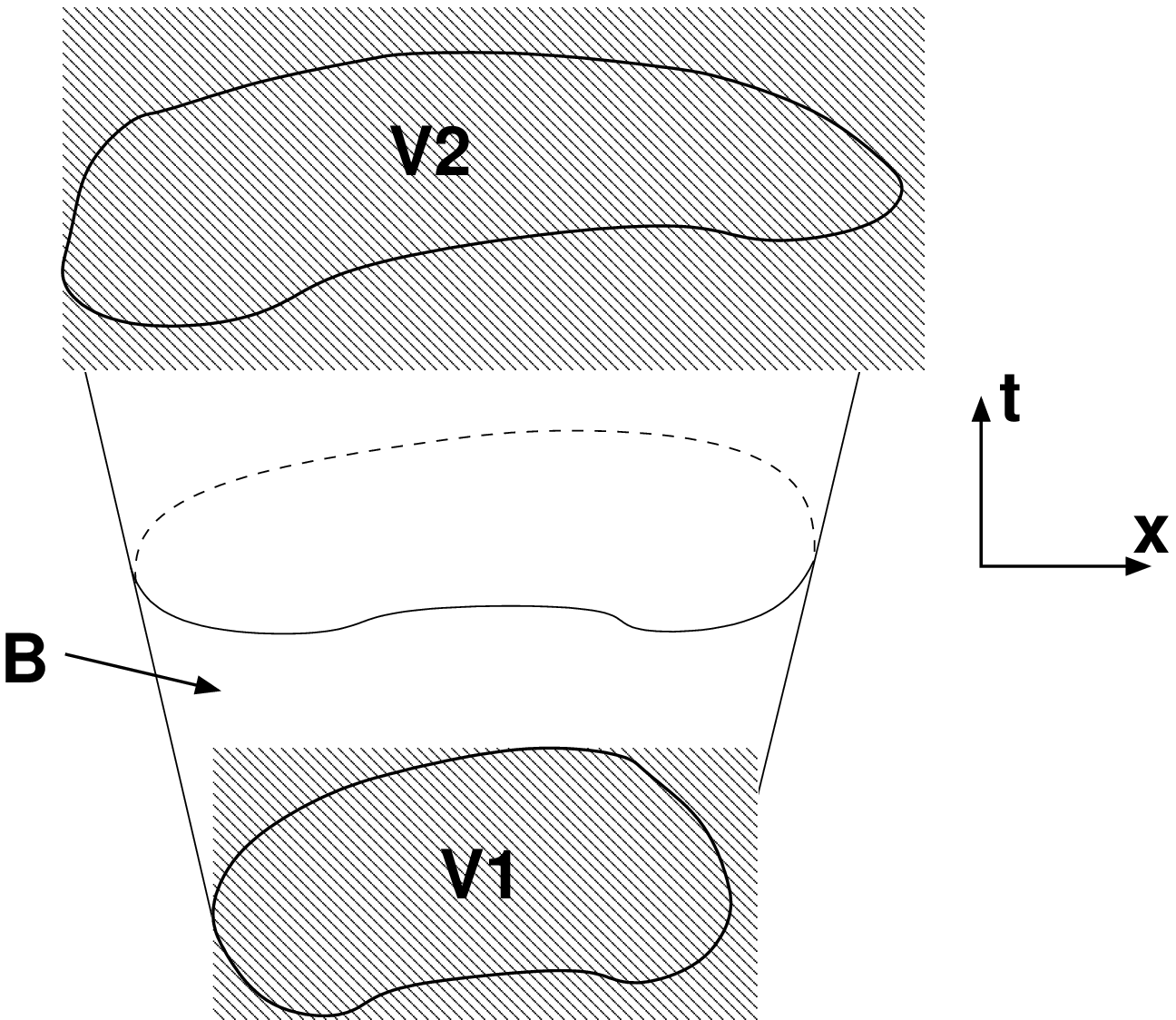}
\end{centre}
Let $V_1$ and $V_2$ be two space-like regions of spacetime joined by a
boundary $B$ which has one space-like and one time-like direction.  Then
using Stokes' theorem we see
$$\int_{V_2} \omega = \int_{V_1} \omega + \int_B \omega$$
Thus it makes sense to regard the integral of $\omega$ over a
space-like region as representing a certain amount of conserved ``stuff''
and the integral over a (space$\times$time)-like region should be
regarded as the flux of the ``stuff'' over time.

Note that there is not a unique $j^\mu$ associated to a symmetry
because we can make the modification
$$j^\mu \longrightarrow j^\mu + \d_\nu (a^{\mu\nu} - a^{\nu\mu})$$ for
any $a^{\nu\mu}$.  So, really conserved currents live in
$$(n-1)\hbox{--forms} \over d\[(n-2)\hbox{--forms}\]$$

\subsection{Not--So--Obvious Symmetries}

As the physics governed by a Lagrangian is determined not by the
Lagrangian but its integral we can allow symmetry transformations
which do not preserve $L$ but change it by a derivative term
$\d_\mu(*)$.  These terms will integrate to zero when we calculate the
action. 

Suppose that we have an infinitesimal symmetry such that
$$\delta L = \d_\mu (K^\mu)$$ 
Then, if we exactly copy the analysis from the previous section we get
a current $j^\mu - k^\mu$ which satisfies the equation
$$\d_\mu (j^\mu - k^\mu) = 0 \in \V/EL$$

To illustrate this, consider the real scalar field with Lagrangian
$$\d_\mu\varphi\d^\mu\varphi + m^2\varphi^2$$
Hence
\begin{eqnarray*}
\delta L &=& \d_\mu \delta \varphi\d^\mu\varphi + \d_\mu\varphi
\d^\mu \delta \varphi + m^2\delta \varphi \varphi + m^2\varphi
\delta \varphi \\
&=& 2\d_\mu \(\delta \varphi\d^\mu\varphi\) -
2\delta \varphi\d_\mu\d^\mu \varphi
+ 2m^2\varphi \delta \varphi \\
&=& \delta \varphi \cdot (2m^2\varphi - 2\d_\mu\d^\mu\varphi)
+ \d_\mu \(2\delta \varphi \d^\mu\varphi\)
\end{eqnarray*}
Note we can see the Euler--Lagrange equations in the final expression for
$\delta L$.  Consider the transformation
$$\varphi\longrightarrow e^{\theta\d_\nu}\varphi$$
whose infinitesimal generator is
$$\delta \varphi = \d_\nu\varphi$$
Substituting this into the final expression for $\delta L$ we easily check
that
$$\delta L=\d_\nu L = \d_\mu\( \delta^\mu_\nu L\)$$
This shows that the above transformation is a symmetry of the physics
described by the Lagrangian.  It also shows that $K^\mu_\nu =
\delta^\mu_\nu L$ The conserved current is therefore
$$j^\mu_\nu = 2\d_\nu\varphi \d^\mu\varphi - \delta^\mu_\nu L$$

This current is called the {\bf energy--momentum tensor}.  As we were
assuming all our Lagrangians are spacetime translation invariant, all
our (classical) theories will have an energy--momentum tensor.
Often the lower index of this current is raised, the resulting current
is
$$T^{\mu\nu} = 2\d^\mu\varphi\d^\nu\varphi - g^{\mu\nu}L$$

\subsection{The Electromagnetic Field}

The space of fields, $\Phi$, has linearly independent elements denoted
by $A^\mu$ such that $A^{\mu\star} = A^\mu$.  The {\bf field strength}
is defined to be
$$F_{\mu\nu} = \d_\nu A_\mu - \d_\mu A_\nu$$
We suppose that there are other fields, denoted by $J^\mu$, which
interact with the electromagnetic field.  In the absense of the
electromagnetic field these $J^\mu$ are assumed to be governed by the
Lagrangian $L_J$.  The Lagrangian for the whole system is
$$L = -{1\over 4} F^{\mu\nu}F_{\mu\nu} - J^\mu A_\mu + L_J$$
The Euler--Lagrange equations obtained by varying $A_\mu$ are
$$\d_\mu F^{\mu\nu} = J^\nu$$
Let us compare this to the vector field form of Maxwell's equations
\begin{eqnarray*}
\nabla\cdot {\mathbf E} &=& \rho \\
\nabla\cdot {\mathbf B} &=& 0 \\
\nabla\times {\mathbf B} - {\d\over\d t}{\mathbf E} &=& {\mathbf j} \\
\nabla\times {\mathbf E} + {\d\over\d t}{\mathbf B} &=& 0
\end{eqnarray*}
Denote the components of the vector ${\mathbf E}$ by $E_x$, $E_y$,
$E_z$ and similarly for the vectors ${\mathbf B}$ and ${\mathbf j}$.
Define
$$F^{\mu\nu} = \(\begin{array}{cccc}
0&-E_x&-E_y&-E_z \\
E_x&0&-B_z&B_y \\
E_y&B_z&0&-B_x \\
E_z&-B_y&B_x&0\end{array}\) \qquad J^\mu = \(\begin{array}{c}
\rho\\j_x\\j_y\\j_z\end{array}\)$$
Then it is easy to see that the first and third of Maxwell's equations
are exactly the Euler--Lagrange equations.
\footnote{
A modern geometrical way to think about Maxwell's equations is to
regard $A$ a a connection on a $\lieU(1)$--bundle.  $F$ is the
curvature of $A$ and the Lagrangian is $L=F\wedge*F$.  Maxwell's
equations then read $dF = 0$ and $d*F = J$.}
The remaining two of
Maxwell's equations follow identically from the fact that
$$F_{\mu\nu} = \d_\nu A_\mu - \d_\mu A_\nu$$

From the Euler--Lagrange equation $\d_\mu F^{\mu\nu} = J^\nu$ it is
clear that
$$\d_\nu J^\nu = 0 \in \V/EL$$
Thus, $J^\nu$ is a conserved current for the theory.  

Introduce a new field $\chi$ such that $\chi^\star = \chi$ and
consider the transformation
$$A_\mu \longrightarrow A_\mu + \theta\d_\mu \chi$$
whose infinitesimal generator is
$$\delta A_\mu = \d_\mu \chi$$
Recall the Lagrangian is of the form
$$L = -{1\over 4} F^{\mu\nu}F_{\mu\nu} + J^\mu A_\mu + L_J$$
It is clear that the term $F^{\mu\nu}F_{\mu\nu}$ is invariant under
the transformation.  What are the conditions required for this
to remain a symmetry of the complete Lagrangian?
$$\delta L = \delta A_\mu (\d_\nu F^{\mu\nu} - J^\mu) + \d_\mu (F^{\mu\nu}\delta A_\nu)
+ \delta J_\mu\times (\cdots) + \d_\mu(\cdots)$$
Substituting $\delta A_\mu = \d_\mu \chi$ gives
$$\delta L = \chi \d_\mu J^\mu + \d_\mu(\cdots)$$
Hence, for the transformation to remain a symmetry we must have that 
$$\chi\d_\mu J^\mu = \d_\mu(\cdots) \quad \hbox{for any $\chi$}$$
This clearly means that
$$\d_\mu J^\mu = 0$$
Note that this equation holds in $\V$ not just in $\V/EL$.  We haven't
shown that $J^\mu$ is a conserved current of the system what we have
shown is that this equation is forced to hold if we want the current
$J^\mu$ to interact with the electromagnetic field as $A_\mu J^\mu$.
We were able to deduce this because $\chi$ was arbitrary --- we have
just seen our first example of a ``local symmetry''.

\subsection{Converting Classical Field Theory to Homological Algebra}

In this section we show how to reduce many of the computation we have
performed into questions in homological algebra.

We give a new description of the module $\Omega^1(V)$.  Let $R$ be a
ring (usually $\CCC[\d_\mu]$) and $V$ an $R$--algebra (usually the
space of Lagrangians).  Suppose that $M$ is any bimodule over $V$.  A
{\bf derivation} $\delta:V\rightarrow M$ is an $R$--linear map which
satisfies the Leibnitz rule
$$\delta(ab) = (\delta a)b + a(\delta b)$$
The {\bf universal derivation module} $\Omega^1(V)$ is
\begin{enumerate}
\item A $V$--bimodule $\Omega^1(V)$
\item A derivation $\delta:V\rightarrow \Omega^1(V)$
\item The following universal property

\emph{For any derivation $d:V\rightarrow M$ there is a unique
$V$--homomorphism $f:\Omega^1(V)\rightarrow M$ such that $d =
f\circ \delta$.}
\end{enumerate}
We define also
$$\Omega^0(V) = V \qquad \Omega^i(V) = \bigwedge\nolimits^i\Omega^1(V)
\qquad \Omega(V) = \bigoplus_i\Omega^i(V)$$
$\Omega(V)$ is $\ZZZ$--graded by assigning grade $i$ to $\Omega^i(V)$.
We denote the grade of a homogeneous element $a$ by $|a|$.  The
derivation $\delta$ extends uniquely to a derivation
$\delta :\Omega(V)\rightarrow\Omega(V)$ such that
$$\delta (ab) = (\delta a)b + (-1)^{|a|}a(\delta b) \qquad \delta ^2=0$$
Define a derivation $\d$ on $\bigwedge(\RRR^n)\tensor\Omega(V)$ by
$$\d(u\tensor v) = \sum_\mu(dx^\mu \wedge u)\tensor\d_\mu v $$
We can now form the following double complex
\begin{diagram}
\wedge^n(\RRR^n)\tensor\Omega^0(V) & \rTo^\delta &
\wedge^n(\RRR^n)\tensor\Omega^1(V) & \rTo^\delta &
\wedge^n(\RRR^n)\tensor\Omega^2(V) & \rTo^\delta & \cdots \\
\uTo>{\d} && \uTo>{\d} && \uTo>{\d} && \\
\wedge^{n-1}(\RRR^n)\tensor\Omega^0(V) & \rTo^\delta &
\wedge^{n-1}(\RRR^n)\tensor\Omega^1(V) & \rTo^\delta &
\wedge^{n-1}(\RRR^n)\tensor\Omega^2(V) & \rTo^\delta & \cdots \\
\uTo>{\d} && \uTo>{\d} && \uTo>{\d} && \\
\vdots && \vdots && \vdots && \ddots
\end{diagram}

It is easy to see that $\delta^2 = \d^2 = 0$ and that $\delta$ and
$\d$ commute.  One can show that all the rows are exact except
possibly at the left hand edge.  Also, all columns are exact except
possibly at the top and left edges.  This means that almost all the
homology groups are zero, however the non-zero ones are usually
interesting things like Lagrangians, currents, ...

\eject\section{Feynman Path Integrals}

In this section we will attempt to define what we mean by
$$\int \exp\( i\int L[\varphi] d^nx\) \D\varphi$$
Note that, even if we manage to make sense of this infinite
dimensional integral, this is only a single real number.  A sensible
description of reality obviously can not be reduced to calculating a
single real number!  So, we actually try to define the integral
$$\int \exp\( i\int L[\varphi] + J\varphi d^nx\) \D\varphi$$
The simplest Lagrangian that illustrates many of the ideas and
problems is
$$L = \d_\mu\varphi \d^\mu\varphi - {m^2 \over 2} \varphi^2 - {\lambda
\over 4!}\varphi^4$$
What do the three terms in this Lagrangian mean?
\begin{itemise}
\item[$\d_\mu\varphi \d^\mu\varphi$] This makes $\varphi(x)$ depend on
$\varphi(y)$ for $x$ near to $y$.
\item[${m^2 \over 2}\varphi^2$] This is a ``mass term''.  It isn't
necessary but it will remove some IR divergences.
\item[${\lambda \over 4!}\varphi^4$] This is a non-quadratic term that
leads to interesting interactions.
\end{itemise}
To evaluate the Feynman path integral we will expand the integrand as
a formal power series in the constant $\lambda$.  This reduced the
problem to computing integrals of the form
$$\int e^{i(\hbox{quadratic in $\varphi$})}\times\prod\(
\int\hbox{polynomial in $\varphi$ and derivatives}\) \D\varphi$$
We can work a way to define these sorts of integral by thinking about
their finite dimensional analogues.
\subsection{Finite Dimensional Integrals}
We will start with a very easy integral and then generalise until we
get to the type of integrals we actually want to evaluate.

Consider the integral
$$\int e^{-(x,x)}d^nx$$
This is well known and easy to evaluate.  The result is $\sqrt{\pi}^n$.

Consider the integral
$$\int e^{i(x,Ax)}d^nx$$
where $A$ is a symmetric matrix with positive definite imaginary
part.  The condition about the imaginary part means that the integral
converges.  By a change of variables we find that the integral is
$$\sqrt{\pi}^n \over \sqrt{\det(A/i)}$$

Consider the integral
$$\int e^{i(x,Ax) + i(j,x)}d^nx$$
where $A$ is as above.  By completing the square we easily see that
the integral is
$$e^{-i(A^{-1}j,j)/4} {\sqrt{\pi}^n \over \sqrt{\det(A/i)}}$$

\subsection{The Free Field Case}

Ignore for now the $\varphi^4$ term in the Lagrangian.  This means
that we are dealing with a ``free field''.  We will use
the results of the last section to ``define'' the Feynman path
integral.  Firstly we need to bring the Feynman integral into a form
covered by the integrals of the previous section.
\begin{eqnarray*}
I[J] &=& \int \exp\( i\int {1\over 2} \d_\mu\varphi \d^\mu\varphi -
{1\over 2}m^2\varphi^2 d^nx + i\int J\varphi d^nx\) \D\varphi\\
&=& \int \exp\( i\int -{1\over 2} \varphi \d_\mu\d^\mu\varphi -
{1\over 2}m^2\varphi^2 d^nx + i\int J\varphi d^nx\) \D\varphi\\
&=& \int \exp\( i\varphi\times -{1\over 2}\[\d_\mu\d^\mu + m^2\]
\varphi + i\int J\varphi d^nx\) \D\varphi
\end{eqnarray*}
This now looks like the integral
$$\int e^{i(x,Ax) + i(j,x)}d^nx$$
if we define $A$ to be the operator
$$-{1\over 2}\[\d_\mu\d^\mu + m^2\]$$
and the inner product $(\varphi,\phi)$ to be
$$\int \varphi\phi d^nx$$
Hence we try to define
$$I[J] := \exp\[ -i(J,A^{-1}J)/4\] \times
{\sqrt{\pi}^{\dim(C^\infty(\RRR^n))} \over \sqrt{\det(A/i)}}$$
There are many problems with this definition
\begin{itemise}\parskip 0pt
\item The inverse of $A$ does not exist.
\item The space of functions is infinite dimensional so we get
$\sqrt{\pi}^\infty$.
\item The determinant of $A$ isn't obviously defined.
\end{itemise}
We can circumvent the last two by looking at the ratio
\footnote{This could be thought of as similar to the fact that ${dy
\over dx}$ exists but defining $dx$ and $dy$ alone requires more care.}
$${I[J] \over I[0]} = \exp\[ -i(J,A^{-1}J) /4 \]$$
Unfortunately, $A^{-1}$ still does not exist.  We can get around this
problem by defining $A^{-1}J$ in a distributional sense
$$A^{-1}J := \int \Delta(x-y) J(y) d^ny$$
where $\Delta$ is a {\bf fundamental solution} for the operator $A$,
i.e. $\Delta$ satisfies
$$A\Delta(x) = \delta(x)$$
Note that $\Delta$ is not necessarily unique --- we will deal with
this problem later.

Finally, we can make the following definition
$${I[J] \over I[0]} := \exp\[ -i\(\int J(x)\Delta(x-y)J(y)d^nxd^ny\)/4 \]$$

\subsection{Free Field Green's Functions}

Even though we have managed to define our way out of trouble for the
free field case we will illustrate here the ideas which turn up when
in the non-free case.

Imagine expanding the exponential in the definition of $I[J]/I[0]$ as
a power series
$${I[J] \over I[0]} = \sum_N \int i^N {G_N(x_1,\dots,x_N) \over N!}
J(x_1)\cdots J(x_N) d^nx_1\cdots d^nx_N$$
The functions $G_N(x_1,\dots,x_N)$ are called the {\bf Green's
functions} for the system.  Knowing all of them is equivalent to
having full peturbative information about the system.  It is possible
that they will not capture some non-peturbative information.  In the
case we are dealing with the first few Green's functions are easy to
work out
\begin{eqnarray*}
G_0 &=& 1 \\
G_2 &=& \Delta(x_1-x_2) \\
G_4 &=& \Delta(x_1-x_2)\Delta(x_3-x_4) + 
\Delta(x_1-x_3)\Delta(x_2-x_4) + \Delta(x_1-x_4)\Delta(x_2-x_3)
\end{eqnarray*}
These can be represented in a diagrammatic form.  We have a vertex for
each argument of the function $G_N$ and join vertices $i$ and $j$ by
an edge if the term $\Delta(x_i - x_j)$ occurs in the expression for
$G_N$.  For example
\begin{eqnarray*}
G_2 &=&
\parbox{10mm}{\begin{fmfgraph}(7,7)\fmfleftn{i}{2}\fmf{plain}{i1,i2}
\fmfdotn{i}{2}\end{fmfgraph}}\\ \\
G_4 &=&
\parbox{10mm}{\begin{fmfgraph}(7,7)\fmfleftn{i}{2}\fmfrightn{o}{2}
\fmf{plain}{i1,o1}\fmf{plain}{i2,o2}\fmfdotn{i}{2}\fmfdotn{o}{2}
\end{fmfgraph}} +\quad
\parbox{10mm}{\begin{fmfgraph}(7,7)\fmfleftn{i}{2}\fmfrightn{o}{2}
\fmf{plain}{i1,o2}\fmf{plain}{i2,o1}\fmfdotn{i}{2}\fmfdotn{o}{2}
\end{fmfgraph}} +\quad
\parbox{10mm}{\begin{fmfgraph}(7,7)\fmfleftn{i}{2}\fmfrightn{o}{2}
\fmf{plain}{i1,i2}\fmf{plain}{o1,o2}\fmfdotn{i}{2}\fmfdotn{o}{2}
\end{fmfgraph}}
\end{eqnarray*}
These diagrams are called {\bf Feynman diagrams}.  Later on they will be
interpreted as representing a particle propagating from the position
represented by a vertex to the position represented by its
neighbouring vertex.

\subsection{The Non-Free Case}

We now try to evaluate the Feynman path integral
$$\int \exp\[ i\int L + J\varphi + {\lambda \over 4!}\varphi^4 d^nx \]
\D\varphi$$
where $L$ is the free field Lagrangian.  We can expand the exponential
terms involving $J$ and $\lambda$ as before to get
$$\int e^{i\int L d^nx} \times \sum_k {i^k \over k!} \( \int J\varphi +
{\lambda \over 4!}\varphi^4 d^nx \)^k \D\varphi$$
This means that we have to deal with terms of the form
$$\int e^{i\int Ld^nx} \times \(\int \varphi(x_1)^{n_1} J(x_1)
d^nx_1\) \(\int \varphi(x_2)^{n_2} J(x_2)d^nx_2\)\cdots \D\varphi$$
We can re-write this integral in the form
$$\int J(x_1)\cdots J(x_N) \times \[\int
e^{i\int Ld^nx} \varphi(x_1)^{n_1}\cdots\varphi(x_N)^{n_N}
\D\varphi\] d^nx_1\cdots d^nx_N$$
The path integral on the inside of the brackets looks very similar to
the integral that turned up in the free field case.  Thus, we could
expect that its value is proportional to
$$G_\bullet (x_1,x_1,\dots,x_2,\dots,\dots)$$
Unfortunately, the Green's function is extremely singular when two of
its arguments become equal.  We will see how to deal with this later.

\eject\section{$0$-Dimensional QFT}

To illustrate some of the ideas we will consider quantum field theory
when spacetime is $0$ dimensional.  In this case the integral over
spacetime is trivial and the path integral over all functions is just
an integral over the real line.

Take the Lagrangian to be
$$L = -{1\over 2} \varphi^2 - {\lambda\over 4!} \varphi^4$$
Note, we no longer have derivative terms because they do not make
sense in $0$ dimensions.

Ignoring the factor of $i$, the Feynman path integral for this
Lagrangian with additional current $j$ is given by
$$\Z[\lambda,j] := \int \exp\( -{1\over 2}\varphi^2 - {\lambda \over
4!}\varphi^4 + j\varphi\) d\varphi$$
If we make the change of variables $\varphi \rightarrow
\varphi\lambda^{-1/4}$ it is easy to see that, considered as a
function of $\lambda$, this integral converges to an analytic function
for $\lambda\ne 0$.  It is also easy to see that there is an essential
singularity and a $4^{th}$--order branch point at $\lambda=0$.

Let us now pretend that we didn't know this function was really nasty
at the origin and attempted to expand it as a power series in
$\lambda$.  Write
$$\exp\(-{\lambda\over 4!}\varphi^4\) = \sum_m {(-\lambda)^m \varphi^{4m}
\over (4!)^m m!} \qquad \exp\(j\varphi\) = \sum_{2k} {j^{2k}
\varphi^{2k} \over (2k)!}$$
Substituting these into the integral for $\Z[\lambda,j]$ we get
$$\Z[\lambda,j] = \int \sum {(-\lambda)^m j^{2k} \over (4!)^m m!
(2k)!} \varphi^{4m+2k} \exp\(-{1\over 2}\varphi^2\)$$
As before, it is very easy to evaluate integrals of the form
$$\int \varphi^{2n} \exp\(-{1\over 2} \varphi^2\) d\varphi$$
A simple induction gives
$${(2n)! \over n!2^n} \times \int \exp\( -{1\over 2}\varphi^2\)
d\varphi$$
Rather than use this as our answer we want to give a graphical
interpretation to the factor $(2n)!/n!2^n$ occuring in this integral.
It is easy to prove by induction that this is exactly the number of
ways of connecting $2n$ dots in pairs.  This observation will be the
origin of the Feynman diagrams in $0$ dimensional QFT.

We can now evaluate the first few terms in the expansion of
$\Z[\lambda,j]$
$${\Z[\lambda,j] \over \sqrt{2\pi}} = 1 - {1 \over 8}\lambda + {5\cdot
7 \over 2^7 \cdot 3} \lambda^2 - {5\cdot 7\cdot 11 \over 2^{10} \cdot
3}\lambda^3 + \cdots + {1\over 2}j^2 + \cdots$$

Let us now give a graphical interpretation to the numbers that turn up
in this expression.  Recall that
$$\Z[\lambda,j] = \sum \int {(-\lambda)^m j^{2k} \over (4!)^m m!
(2k)!} \varphi^{4m+2k} \exp\(-{1\over 2}\varphi^2\)$$
We could also write this in the form
$$\Z[\lambda,j] = \sum \int \overbrace{{{(-\lambda)\varphi^4 \over 4!} \times
\cdots \times {(-\lambda)\varphi^4 \over 4!} \over m!}}^m \cdot
\overbrace{{j\varphi \times\cdots \times j\varphi \over (2k)!}}^{2k} \cdot
\exp\(-{1\over 2}\varphi^2\) d\varphi$$
We could just naively write down a graph with $4m+2k$ vertices and try
to join them up in pairs.  This would give us the correct answer but
there is a much nicer way to do this.

Notice that some of the factors
of $\varphi$ come from the $\varphi^4$ term and some from the
$j\varphi$ term.  To graphically keep track of this introduce two
different types of vertices.
\begin{itemise}\parskip 0pt
\item $m$ vertices with valence $4$ which correspond to the $m$
$\varphi^4$ terms in the integral.
\item $2k$ vertices with valence $1$ which correspond to the $2k$
$j\varphi$ terms in the integral.
\end{itemise}
We now join the vertices up in all possible ways.\footnote{
Note this is identical to taking the graph on $4m+2k$ vertices and
joining all vertices in pairs and then identifying the vertices which
come from the same $\varphi^4$ term or the same $j\varphi$ term.
However in the current form we know how many factors of $j$ there are
in the integral which we would not know if all the vertices looked
identical.}  It is important to
remember that we consider the vertices and edges as labeled
(i.e. distinguishable).  For example, if we only have one vertex of
valence $4$ then joining edge $1$ to $2$ and edge $3$ to $4$ is
{\bf different} to joining edge $1$ to $3$ and edge $2$ to $4$.

So, we can replace our integral by a sum over graphs provided we know
how to keep track of all the $\lambda$'s, $j$'s and factorials.
Consider the action of the group which permutes vertices and edges ---
this has order $(4!)^m m! (2k)!$.  Notice that this is exactly the
combinatorial factor that should be attached to each graph.  Hence, by
the Orbit--Stabiliser theorem we can replace the sum over all graphs
by the sum over isomorphism classes of graphs weighted by
$1/\Aut(g)$.  This leads to the following graphical evaluation of the
integral:  For each isomorphism class of graphs associate factors of
\begin{itemise}\parskip 0pt
\item $(-\lambda)$ for each vertex of valence $4$.
\item $j$ for each vertex of valence $1$.
\item $1/\Aut(g)$.
\end{itemise}
The integral is then the sum of these terms.  To illustrate this we
consider the first few terms that one obtains from this graphical expansion
\begin{eqnarray*}
\hbox{Empty graph} & \longrightarrow & 1 \\
\parbox{20mm}{
\begin{fmfgraph}(15,15)\fmfkeep{eight}
\fmfleft{i}
\fmfright{o}
\fmf{phantom}{i,v,o}
\fmf{plain}{v,v}
\fmf{plain,left=90}{v,v}
\fmfdot{v}
\end{fmfgraph}} & \longrightarrow & -{\lambda \over 8} \\
\parbox{25mm}{\fmfreuse{eight}\hskip-1cm\fmfreuse{eight}} & \longrightarrow &
{\lambda^2 \over 128} \\
\parbox{20mm}{
\begin{fmfgraph}(15,15)
\fmfleft{i}
\fmfright{o}
\fmf{plain,left=1}{i,v,i}
\fmf{plain,left=1}{o,w,o}
\fmf{plain,left=1}{v,w,v}
\fmfdot{v}
\fmfdot{w}
\end{fmfgraph}} & \longrightarrow &
{\lambda^2 \over 16} \\
\parbox{20mm}{\begin{fmfgraph}(15,15)
\fmfleft{i}
\fmfright{o}
\fmf{plain,left=0.3}{i,o,i}
\fmf{plain,left=.7}{i,o,i}
\fmfdot{i}
\fmfdot{o}
\end{fmfgraph}} & \longrightarrow &
{\lambda^2 \over 48}
\end{eqnarray*}
Putting these all together gives
$$1 - {1 \over 8}\lambda + {5\cdot 7 \over 2^7 \cdot 3}\lambda^2
+\cdots$$
which is exactly what was obtained before.

It is now time to remember that the expansion we performed was centred
at an essential singularity and so is not valid as a power series.
What we will show now is that it is an asymptotic expansion for the
integral.  For simplicity we will do this for the function
$\Z[\lambda,0]$ --- i.e. we ignore all the $j$ terms.

Recall that
$$\left|\exp(x) - \sum_{n=0}^k {x^n \over n!}\right| \le {x^{k+1}
\over (k+1)!}$$
This means that the error in computing the integral for
$\Z[\lambda,0]$ by taking the first $k$ terms of the series obtained
above is bounded by $C\lambda^{k+1}$.  $C$ is a constant that depends
on $k$ but not on $\lambda$.  This shows that the series is an
asymptotic expansion for the integral.\footnote{An asymptotic
expansion is something that becomes more accurate as
$\lambda\rightarrow 0$ with $k$ fixed.  A power series is something
which becomes more accurate as $k\rightarrow\infty$ with $\lambda$ fixed.}

What is the maximum accuracy to which we can compute $\Z[\lambda,0]$
with the asymptotic expansion?  To determine this we need to find the
smallest term in the asymptotic expansion as this will give the
error.  The ratio of consecutive terms is roughly ${2\lambda n \over
3}$.  When this ratio is $1$ is the point where the terms are
smallest.  So we find that we should take
$$k \approx {3\over 2\lambda}$$
If we do this then the error turns out to be in the region of
$e^{-{3\over 2\lambda}}$.

\subsection{Borel Summation}

The series which we computed for $\Z[\lambda,0]$ is of the form
$$\sum_n a_n x^n \quad \hbox{with $a_n \approx Cn!$}$$
There are known ways to sum series of this form, one of which is known
as Borel summation.

The trick is to notice that
$$\int_0^\infty e^{-t/x}t^n dt = x^{n+1} n!$$
Hence
$$\int_0^\infty e^{-t/x} \( \sum_n {a_n \over n!} t^n\) {dt
\over x} = \sum_n a_n x^n$$
If the function
$$g(t) := \sum_n {a_n \over n!} t^n$$
extends to all $t>0$ with sufficiently slow growth then we will be
able to compute the integral and so define the ``sum'' of the series.

There are, however, a couple of problems with this method in quantum
field theory
\begin{itemise}\parskip 0pt
\item It does not pick up non-peturbative effects.
\item $g(t)$ looks like it might have singularities on $t>0$.
\end{itemise}

\subsection{Other Graph Sums}

We give two tricks which allow the graph sum to be simplified
slightly.

We have seen that $\Z[\lambda,j]$ is given by a sum over isomorphism
classes of graphs.  Consider now the function\footnote{Warning: Some
books use the notation $\W$ and $\Z$ the other way around}
$$\W[\lambda,j] := \log \Z[\lambda,j]$$
We claim that this can be evaluated by a graph sum where this time the
sum is over isomorphism classes of connected graphs.  To show this we
will compute the exponential of the graph sum for $\W$ and show that
it is the same as the graph sum for $\Z$.  Write $a[G]$ for the
factors of $\lambda$ and $j$ that are attached to the isomorphism
class of the graph $G$.  Then
\begin{eqnarray*}
\exp\( \sum_{\hbox{\footnotesize conn}} {a[G] \over \Aut(G)} \) &=&
\prod_{\hbox{\footnotesize conn}} \exp\( {a[G] \over \Aut(G)} \) \\
&=& \prod_{\hbox{\footnotesize conn}} \sum_{n=0}^\infty {a[G]^n \over n!
\Aut(G)^n} \\
&=& \prod_{\hbox{\footnotesize conn}} \sum_{n=0}^\infty {a[nG] \over \Aut(nG)}
\\
&=& \Z[\lambda,j]
\end{eqnarray*}

A second simplification can be obtained by remembering that we wanted
the ratio $\Z[\lambda,j] \over \Z[\lambda,0]$ rather than the function
itself.  It is easy to see that this is obtained by exponentiating the
sum over all connected graphs with at least one valence $1$ vertex.
This is sometimes called removing the ``vacuum bubbles''.

\subsection{The Classical Field}

Using the idea that dominant terms in quantum physics are the ones
close to the classical solutions we could try to define a field 
$\varphi_{cl}$ for a given $j$ which is the ``average'' of the quantum
fields.  We might hope that this satisfies the Euler--Lagrange
equations.

Define
$$\varphi_{cl} = {\int \varphi \exp\( -{1\over 2}\varphi^2 - {\lambda
\over 4!}\varphi^4 + j\varphi\) d\varphi \over \int \exp\( -{1 \over
2}\varphi^2 - {\lambda \over 4!}\varphi^4 + j\varphi\) d\varphi}$$
In terms of the functions $\Z$ and $\W$ it can be given by
$${{d\over dj}\Z[\lambda,j] \over \Z[\lambda,j]} = {d\over dj}
\W[\lambda,j]$$
Thus, there is an obvious graphical way to compute this function.  We
do a sum over connected graphs with the numerical factors changed in
the obvious way to take into account the derivative with respect to
$j$.  So, to an isomorphism class of graphs we attach to factor
\begin{itemise}\parskip 0pt
\item $(-\lambda)^n$ if there are $n$ vertices of valence $4$.
\item $m\times j^{m-1}$ if there are $m$ vertices of valence $1$.
\item $1/\Aut(g)$.
\end{itemise}
For example
\begin{eqnarray*}
\parbox{20mm}{\begin{fmfgraph}(15,15)
\fmfleft{i}
\fmfright{o}
\fmf{plain}{i,o}
\fmfdot{i}
\fmfdot{o}
\end{fmfgraph}}
&\longrightarrow& j \\
\parbox{17.5mm}{\begin{fmfgraph}(10,10)
\fmfleftn{i}{2}
\fmfrightn{o}{2}
\fmf{plain}{i1,v,o1}
\fmf{plain}{i2,v,o2}
\fmfdotn{i}{2}
\fmfdotn{o}{2}
\fmfdot{v}
\end{fmfgraph}}
&\longrightarrow& -{\lambda j^3 \over 3!} \\
\parbox{20mm}{\begin{fmfgraph}(15,15)
\fmfleft{i}
\fmfright{o}
\fmf{plain}{i,v,v,o}
\fmfdot{i}
\fmfdot{o}
\fmfdot{v}
\end{fmfgraph}}
&\longrightarrow& -{\lambda j \over 2}
\end{eqnarray*}
It is now easy to check that the classical field $\varphi_{cl}$ does
not satisfy the Euler--Lagrange equations.  However, if we slightly
modify the definition we can obtain something which is a solution to
the Euler--Lagrange equations.

Define $\varphi_t$ to be the sum for $\varphi_{cl}$ except that we
keep only the graphs which are trees (i.e. have no loops).  It is now
possible to show that this function satisfies the Euler--Lagrange
equations.  To do this we introduce a new variable $\hbar$ into the
integral
$$\Z[\lambda,j,\hbar] := {1 \over \sqrt{\hbar}}\int \exp\(\[-{1\over
2}\varphi - {\lambda \over 4!}\varphi^4 + j\varphi\] / \hbar\)
d\varphi$$
If we change variables by $\varphi \rightarrow \sqrt{\hbar} \varphi$
it is easy to see that the factor of $\hbar$ that occurs attached to a
graph with $v_1$ vertices of valence $1$ and $v_4$ vertices of valence
$4$ is $\hbar^{v_4 - v_1/2}$.  As before we can define
$\W[\lambda,j,\hbar]$ to be its logarithm.  It is easy to see that
this function is given by a sum over connected graphs with the same
numerical factors as above.  We can define a new classical field (now
depending on $\hbar$) by
$$\varphi_{cl}(\lambda,j,\hbar) := \hbar {d\over dj}
\W[\lambda,j,\hbar]$$
We see that this is given by the same graph sum as the previous
classical field except that we have the additional factor of
$\hbar^{1 + v_4 - v_1/2}$ attached to each graph.  Note that the
exponent of this is exactly $1-\chi(G)$ which is the number of loops
in the graph $G$.  So the function $\varphi_t$ is the constant term of
this graph sum in $\hbar$.  Alternatively, we can think of recovering
the function $\varphi_t$ from $\varphi_{cl}$ by taking the limit
$\hbar \rightarrow 0$.

As we take the limit $\hbar\rightarrow 0$ we see that the exponential
in the definition for $\varphi_{cl}$ becomes concentrated around the
value of $\varphi$ which maximises the exponent.  Assuming that this
value is unique we see that $\varphi_{t}$ is the value of $\varphi$
which maximises the exponent.  However, the Euler--Lagrange equations
are exactly the equations which such a maximum will satisfy.  Hence
$\varphi_t$ satisfies the Euler--Lagrange equations.  If the maximum
was not unique then $\varphi_t$ would be some linear combination of
the possible $\varphi$ which maximised the exponent.  However all of
these would individually satisfy the Euler--Lagrange equations and
hence the linear combination would too.

\subsection{The Effective Action}

It is still unsatisfactory that we have a field called the {\bf classical
field} and no classical equations of motion that it satisfies.  If we
could find a new Lagrangian such that the old $\varphi_{cl}$ was equal
to the new $\varphi_t$ then $\varphi_{cl}$ would satisfy the new
Euler--Lagrange equations.  To do this we recall that any connected
graph can be written in the following ``tree form''

$$\begin{fmfgraph}(45,45)
\fmfleftn{i}{4}
\fmfrightn{o}{4}
\fmf{plain}{i2,v1,i3}
\fmfblob{.15w}{v1}
\fmf{plain,tension=2}{v1,w,v2}
\fmfblob{.15w}{v2}
\fmfdot{w}
\fmf{plain,tension=0.5}{v2,v3,o2}
\fmfblob{.15w}{v3}
\fmf{plain}{v2,v4}
\fmfdot{v4}
\fmf{plain}{o3,v4,o4}
\end{fmfgraph}$$\vskip -1cm
Where the blobs are subgraphs that can not be disconnected by cutting
a single edge.  These are called {\bf 1 particle irreducible} graphs or
1PI for short.  Note that this decomposition is unique.

Recall that to a term of the form $\varphi^n \over n!$ in the
Lagrangian we would have associated a vertex with valence $n$.  In
order to define the new Lagrangian we are trying to regard the 1PI
blobs as vertices.  It is therefore sensible to associate factors like
$\varphi^n$ to 1PI diagrams with valence $n$.  This turns out to
be the correct idea.  To any 1PI diagram attach the following factors
\begin{itemise}\parskip 0pt
\item $(-\lambda)$ for each vertex.
\item $\varphi$ for each unused edge.
\item $1/\Aut(G)$
\end{itemise}
The effective action is then defined to be
$$\Gamma[\varphi] := -{1\over 2}\varphi^2 + \sum_{\hbox{\footnotesize 1PI}}
\hbox{Attached factors}$$ 
To illustrate this we compute the corrections up to the two vertex
level (the unused edges are marked with dotted lines)
\begin{centre}
\begin{minipage}{2.5in}
\begin{eqnarray*}
\parbox{12mm}{\begin{fmfgraph}(7,7)
\fmfleft{i}
\fmfright{o}
\fmftop{t}
\fmfbottom{b}
\fmf{dots}{i,v,o}
\fmf{dots}{t,v,b}
\fmfdot{v}
\end{fmfgraph}}
& \longrightarrow & -\lambda {\varphi^4 \over 4!} \\
\parbox{15mm}{\vskip 5mm\begin{fmfgraph}(10,10)
\fmfleft{i}
\fmfright{b}
\fmf{dots}{i,v,b}
\fmf{plain,tension=.5}{v,v}
\fmfdot{v}
\end{fmfgraph}}
& \longrightarrow & -\lambda {\varphi^2 \over 4} \\
\parbox{15mm}{\begin{fmfgraph}(10,10)
\fmfleft{i}
\fmfright{o}
\fmf{plain,left=1}{i,v,i}
\fmf{plain,left=1}{o,v,o}
\fmfdot{v}
\end{fmfgraph}}
& \longrightarrow & -\lambda {1 \over 8} \\
\parbox{15mm}{\begin{fmfgraph}(10,10)
\fmfleft{i}
\fmfright{o}
\fmf{plain,left=1}{i,o,i}
\fmf{plain,left=0.5}{i,o,i}
\fmfdot{i,o}
\end{fmfgraph}}
& \longrightarrow & \lambda^2 {1 \over 48}
\end{eqnarray*}
\end{minipage}
\begin{minipage}{2.5in}
\begin{eqnarray*}
\parbox{19mm}{\begin{fmfgraph}(15,10)
\fmfleft{i}
\fmfright{o}
\fmf{plain,left=1}{i,v,i}
\fmf{plain,left=1}{o,w,o}
\fmf{plain,left=1}{v,w,v}
\fmfdot{v,w}
\end{fmfgraph}}
& \longrightarrow & \lambda^2 {1 \over 16} \\
\parbox{19mm}{\begin{fmfgraph}(15,10)
\fmfleft{i}
\fmfrightn{o}{2}
\fmf{plain,left=1,tension=0.3}{i,v,i}
\fmf{plain,left=1,tension=0.3}{v,w,v}
\fmf{dots}{o1,w,o2}
\fmfdot{v,w}
\end{fmfgraph}}
& \longrightarrow & \lambda^2 {\varphi^2 \over 8} \\
\parbox{19mm}{\begin{fmfgraph}(15,10)
\fmfleft{i}
\fmfright{o}
\fmf{dots}{i,v}
\fmf{dots}{o,w}
\fmf{plain,left=1,tension=0.2}{v,w,v}
\fmf{plain,tension=0.2}{v,w}
\fmfdot{v,w}
\end{fmfgraph}}
& \longrightarrow & \lambda^2 {\varphi^2 \over 12} \\
\parbox{19mm}{\begin{fmfgraph}(15,10)
\fmfleftn{i}{2}
\fmfrightn{o}{2}
\fmf{dots}{i1,v,i2}
\fmf{dots}{o1,w,o2}
\fmf{plain,left=1,tension=0.3}{v,w,v}
\fmfdot{v,w}
\end{fmfgraph}}
& \longrightarrow & \lambda^2 {\varphi^4 \over 8}
\end{eqnarray*}
\end{minipage}
\end{centre}
The terms that occur with the factor $\varphi^2$ are sometimes called
``mass corrections'' because they alter the mass term of the original
Lagrangian.  The higher order terms in $\varphi$ are ``field strength
corrections'' as they alter the behaviour of how the physical theory
interacts.  The terms which are independent of $\varphi$ can usually
be ignored as they only shift the Lagrangian by a constant which will
not affect the equations of motion.  There is one situation where this
is not true: If we are varying the space-time metric then these
changes are important.  They give rise to the ``cosmological
constant'' that Einstein introduced into general relativity in order
not to predict the universe was expanding.  Einstein was later to
describe this as the worst mistake he ever made.\footnote{Recent
experiments, however, give a non-zero value to the cosmological
constant.  Unfortunately the theoretical prediction for the value of
the constant is about $10^{120}$ times too large.  This is possibly
the worst theoretical prediction ever!}

If we add up the terms calculated above we find that the effective
action (to order $\lambda^2$) is
$$\Gamma[\varphi] = -{1\over 2}\varphi^2 - {\lambda \over 4!}
\varphi^4 - {\lambda \over 8}\varphi^2 - {\lambda \over 8} +
\O(\lambda^2)$$
The Euler--Lagrange equations for this Lagrangian with a source $j$
are
$$-\varphi - {\lambda \over 3!}\varphi^3 - {\lambda \over 4}\varphi +
j = 0$$
It is now easy to check that the computed value for $\varphi_{cl}$ is
a solution to this.  This means that the following equation holds
$$j = \left. -{d\over d\varphi}\Gamma[\varphi]\right|_{\varphi_{cl}}$$
Recall that $\varphi_{cl}$ was defined by the equation
$$\varphi_{cl} = {d \over dj} \W[j]$$
These two equations combined mean that
$$\Gamma[\varphi_{cl}] = \W[j] - j\varphi_{cl} + \hbox{const}$$
Absorbing the constant into the definition of $\Gamma[\varphi]$ shows
that $\W$ and $\Gamma$ are Legendre transforms of each other.  We now
give a graphical proof of this equation.

We know that $\W[j]$ is given by a sum over isomorphism classes of
connected graphs of
$$(-\lambda)^{v_4} j^{v_1} \over \Aut(g)$$
Hence, $j\varphi_{cl}$ which is $j{d \over dj}\W[j]$ is given by the
sum over isomorphism classes of connected graphs of
$$v_1 \times {(-\lambda)^{v_4} j^{v_1} \over \Aut(g)}$$
This sum can be though of as a sum over isomorphism classes of
connected graphs with a marked valence $1$ vertex of
$$(-\lambda)^{v_4} j^{v_1} \over \Aut(g_v)$$
We want to give an expression for $-{1\over 2}\varphi_{cl}^2$ in terms
of connected graphs.  If we simply square the sum for $\varphi_{cl}$
we get a sum over pairs of isomorphism classes of graphs with marked valence
$1$ vertices of
$$(-\lambda)^{v_4+v_4'} j^{v_1+v_1'-2} \over \Aut(g_v)\Aut(g_v')$$
The obvious way to get a connected graph from a pair of connected
graphs with marked vertices is simply to identify the two marked
vertices.  This gives us a connected graph with $v_4+v_4'$ valence $4$
vertices and $v_1+v_1'-2$ valence $1$ vertices and a marked edge.  It
is now easy to see that ${1\over 2}\varphi_{cl}^2$ is given by the sum
over isomorphism classes of graphs with a marked edge of
$$(-\lambda)^{v_4} j^{v_1} \over \Aut(g_e)$$
The factor of ${1 \over 2}$ accounts for the fact that we can
interchange $g$ and $g'$ if they were different and the automorphism
group $\Aut(g_e)$ is twice as large if $g=g'$.  This sum is also equal
to the sum over isomorphism classes of graphs of
$$e \times {(-\lambda)^{v_4} j^{v_1} \over \Aut(g)}$$
where $e$ is the number of edges that when cut would separate the
graph.

Finally, we need an expression for the sum over 1PI graphs in terms of
connected graphs.  The sum that occurs in the expression for
$\Gamma[\varphi_{cl}]$ is the sum over isomorphism classes of 1PI
graphs of
$$(-\lambda)^{v_4} \varphi_{cl}^{v_u} \over \Aut(g)$$
where $v_u$ is the number of unused edges.  $\varphi_{cl}$ is given by
a sum over graphs with marked valence $1$ vertices.  We use these
marked vertices to attach the graphs from $\varphi_{cl}$ to the unused
edges from the 1PI diagram.  This gives a connected graph with marked
1PI piece.  So we get a sum over isomorphism classes of connected
graphs with marked 1PI piece of
$$(-\lambda)^{v_4} j^{v_1} \over \Aut(g_1)$$
This is easily seen to be equal to the sum over isomorphism classes of
connected graphs of
$$v \times {(-\lambda)^{v_4} j^{v_1} \over \Aut(g)}$$
where $v$ is the number of 1PI pieces of the graph $g$.

Consider the sum
$$\Gamma[\varphi_{cl}] - \W[j] + j\varphi_{cl}$$
This is given by the graph sum over all connected graphs of
$$-1 + v_1(g) - e(g) + p(g)$$
If we think of the graph $g$ in its tree form then it is clear
that this sum is zero.  Hence we have a graphical proof of the fact
that the functions $\Gamma[\varphi_{cl}]$ and $\W[j]$ are Legendre
transforms of each other.

\eject\section{Distributions and Propagators}

Let $C_0^\infty(\RRR^n)$ be the space of smooth compactly supported
function on $\RRR^n$.  The dual space to $C_0^\infty(\RRR^n)$ is
called the space of {\bf distributions}.  These are a generalisation
of functions where we can always differentiate.  To see this consider
the distribution defined below for any locally integrable function
$f(x)$
$$M_f:t \mapsto \int_{\RRR^n} f(x)t(x) dx$$
So the space of locally integrable functions is a subset of the space
of distributions.  If we assume that $f(x)$ was differentiable we
would get by integration by parts
$$M_{f'}:t \mapsto \int_{\RRR^n} f'(x)t(x) dx = -\int_{\RRR^n}
f(x)t'(x)$$
We can use this do define the derivative of any distribution as
$${\d \over \d x}D(t) := -D\({\d \over \d x}t\)$$
So we can differentiate distributions, even ones which came from
functions with discontinuities.  If we differentiate the {\it
Heavyside step function}
$$\theta(x) =
\begin{cases}
0 & \hbox{if $x<0$} \\
1 & \hbox{if $x>0$}
\end{cases}
$$
Then it is an easy exercise to check that we get the following
distribution
$$\delta(t) = t(0)$$
which is called the {\bf Dirac delta function}.  This is one of the
most widely used distributions.

It is easy to define the support of a distribution.  If this is done
then the delta function is a distribution which is supported only at
the origin.  It can be shown that any distribution which is supported
at the origin is a linear combination of derivatives of the delta
function.  This fact will be widely used later.

In general it is not possible to define the product of two
distributions.  Later on in this section we will find a sufficient
condition that will guarantee when a product can be sensibly defined.

A natural operation that turns up in Physics is the Fourier
transform.  It is not possible to define the Fourier transform for the
distributions we have defined (basically they can grow too quickly)
however if we use a more restricted class of distributions we will be
able to do this.  However, before this we recall some basic facts
about Fourier transformations.

If we have a function $f(x)$ then physicists usually define the
Fourier transform to be
$$\hat f(y) = \int e^{-i(x,y)}f(x) dx$$
Let $S(\RRR^n)$ be the {\bf Schwartz space of test functions} which
consists of functions on $\RRR^n$ all of whose derivatives decrease
faster than any polynomial.  It is then standard to check that the
Fourier transform is an automorphism of the Schwartz space.  The
following formulae are also easy and standard to prove
\begin{eqnarray*}
\hat{\hat f}(-x) &=& (2\pi)^n f(x) \\
\widehat{\varphi * \psi} &=& \hat{\varphi}\hat{\psi} \\
\widehat{\varphi \psi} &=& (2\pi)^{-n} \hat\varphi * \hat\psi \\
\int \varphi \bar\psi &=& (2\pi)^{-n} \int \hat\varphi \bar{\hat \psi}
\\
\int \hat\varphi \bar\psi &=& \int \varphi \bar{\hat \psi}
\end{eqnarray*}
The last two equations say that the Fourier transform is almost an
isometry of the Schwartz space.

The dual space of the Schwartz space is the space of {\bf tempered
distributions}.  Roughly speaking this means that the distribution is
of at most polynomial growth (there are examples of functions with
non-polynomial growth which define tempered distributions but they can
in some sense be regarded as pathological).

The main property of tempered distributions we want is that it is
possible to define the Fourier transform on them.  Using the last
equation above we define the Fourier transform of a distribution $D$
as
$$\hat D(t) := D(\hat t)$$

\subsection{Euclidean Propagators}

We use Fourier transforms of distributions to find solutions to
differential equations of the form
$$\( -\sum {\d^2 \over \d x_i^2} + m^2\)f = \delta(x)$$
Recall that our original propagator $\Delta(x)$ was defined to be the
distributional solution to Euler--Lagrange equations and these were of
the same form as the equation above.

We can solve these by taking Fourier transforms and solving the
resulting polynomial equation.  One solution is certainly given by
$$\hat f = {1 \over \sum x_i^2 + m^2}$$
To find the solution $f$ we just use the inverse Fourier transform on
this.

Consider the case when $m\ne 0$.  Then it is easy to see that the
solution for $\hat f$ is unique.  As this solution is smooth the
resulting solution for $f$ is rapidly decreasing (because Fourier
transform swaps differentiability with growth behaviour), this means
that the solution will have no IR divergencies.  As an example, in $1$
dimension the solution is easy to compute
$$f(x) = {\pi \over m}e^{-|x| m}$$

When $m=0$ various things are more difficult.  The obvious choice for
$\hat f$ is no-longer smooth and hence its Fourier transform will have
IR divergencies.  However, there is now more than one possible
solution to the equations.  This time we can hope (as before) that
there is a canonical choice for $\hat f$ --- this isn't always true,
however we do have the following result:

{\it If $f$ is a homogeneous distribution on $\RRR^n - 0$ of degree
$a$ and $a\ne -n,-n-1,\dots$ then it extends uniquely to a homogeneous
distribution on $\RRR^n$.}

Suppose that the distribution $f$ is given by integrating against a
homogeneous function $f$.  Then
$$\<f,\varphi\> = \int_{\RRR^n} f\varphi dx = \int_{S^{n-1}}
\int_{r=0}^\infty f(w) r^{a+n-1} \varphi(rw)drdw$$
We try to re-write this integral so that it defines a distribution on
$\RRR^n$ rather than just $\RRR^n-0$.  Define the following
distribution for $a>-1$
$$t^a_+ =
\begin{cases}
t^a & \hbox{if $t>0$} \\
0 & \hbox{if $t<0$}
\end{cases}
$$
This can be extended to a distribution whenever $a\ne -1,-2,\dots$
\footnote{In fact, $t_+^a$ is a meromorphic distribution with poles at
$a=-1,-2,\dots$.  For example, the residue at $a=-1$ is just the delta
function}.  To do this note that the above distribution satisfies the
differential equation
$${d \over dt} t_+^a = at^{a-1}_+$$
With this distribution on functions on $\RRR$ we can define the
operator
$$R_a(\varphi)(x) = \<t^{a+n-1}_+, \varphi(tx)\>$$
This takes a function $\varphi$ and gives a homogeneous function of
degree $-n-a$.  Pick $\psi$ to be a spherically symmetric function
with compact support on $\RRR^n-0$ such that
$$\int_0^\infty \psi(rx) {dr \over r} = 1$$
This can be thought of as a spherical bump function.  Then it is easy
to check that if $\varphi$ is homogeneous of degree $-n-a$ then
$$R_a(\psi\varphi) = \varphi$$
So, it is possible to recover our test function on $\RRR^n$ from one
which is a test function on $\RRR^n-0$ provided that it was
homogeneous.  Looking at the integral which defined the original
distribution we see that
$$\<f,\varphi\> = \<f,\psi R_a(\varphi)\>$$
if $\varphi$ is a test function on $\RRR^n-0$.  However, the right
hand side is defined for test functions on $\RRR^n$ so we take this to
be the extension.

As any distribution can be obtained by differentiating a function we
see that we have done enough to show the result (checking the
extension is unique and homogeneous is easy).

With the above result we now know that the massless propagator in
dimensions $3$ and higher is unique and it isn't hard to show that it
is $(x^2)^{1-n/2}$.

In dimension $1$ there are many solutions ${|x| \over 2} + kx$ but the
constant $k$ can canonically be set to zero by insisting on symmetry
under $x\rightarrow -x$.

In dimension $2$ there are many solutions $2\ln|x| + k$ and this time
there is no way to canonically set the constant to zero.  Note that in
this case not only was the extension non-unique but it wasn't even
homogeneous.

\subsection{Lorentzian Propagators}

The propagators we are really interested in in QFT live in Lorentz
space.  The new problem we have to deal with is that they now have
singularities away from the origin.

{\bf Warning:} Due to the difference in sign conventions in Lorentz
space $(+,-,-,-)$ and Euclidean space $(+,+,+)$ the differential
equations change by a minus sign.  In Lorentzian space we are trying
to solve the following differential equation
$$\(\sum \d_i \d^i + m^2\)f = \delta(x)$$
By taking Fourier transforms we need to find distributional solutions
to the equation
$$(p^2 - m^2)\hat f = 1$$
Before solving this equation let us see how unique the solutions are.
This involves looking for solutions of the equation
$$(p^2-m^2)\hat f = 0$$
These are clearly given by distributions supported on the two
hyperboloids $S_1$ and $S_2$ defined by $p^2=m^2$.
\begin{centre}
\includegraphics[scale=.4]{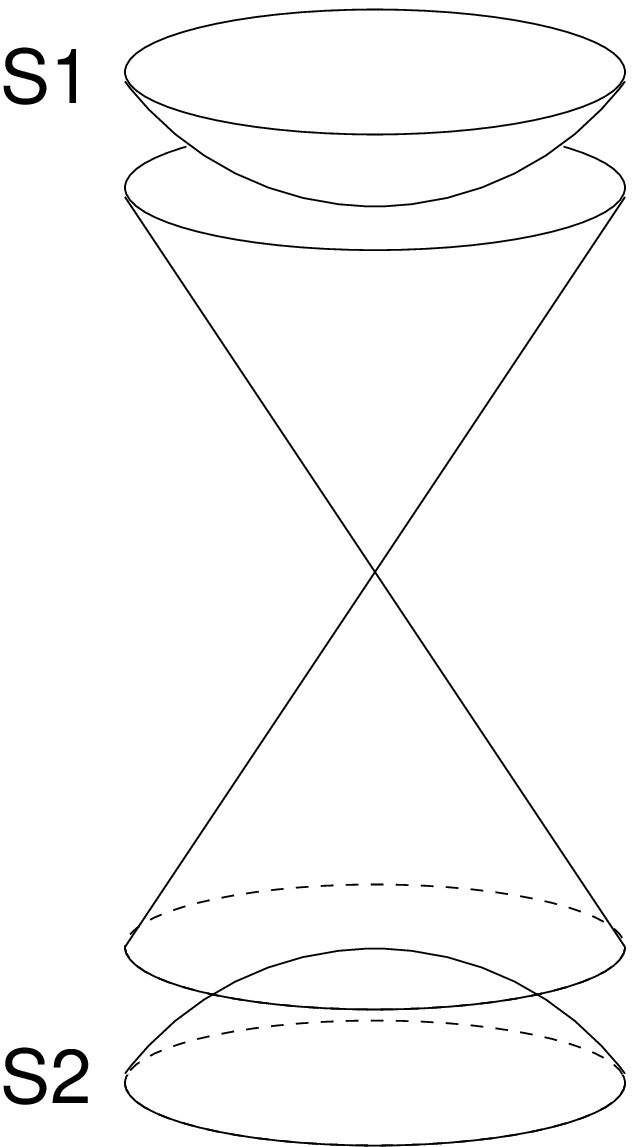}
\end{centre}
There are rather too many of these so we cut down the number by
insisting that all our solutions are invariant under the connected
rotation group of Lorentzian space.  This means that there is now only
a two dimensional set of solutions.

To work out solutions to the original differential equation we need to
work out the inverse Fourier transform of
$$\hat f(p) = {1 \over p^2 - m^2}$$
The integral we are trying to compute is
$$\int {\exp(i(x_0p_0 - x_1p_1 - \cdots)) \over p_0^2-p_1^2-\cdots -
m^2} dp$$
Which has singularities at $p_0 = \pm\sqrt{p_1^2+\cdots + m^2}$.
Regard all the variables as complex rather than real and we can
evaluate the integral as a contour integral.  In the $p_0$--plane we
can use a contour like the following.
\begin{centre}
\includegraphics[scale=.6]{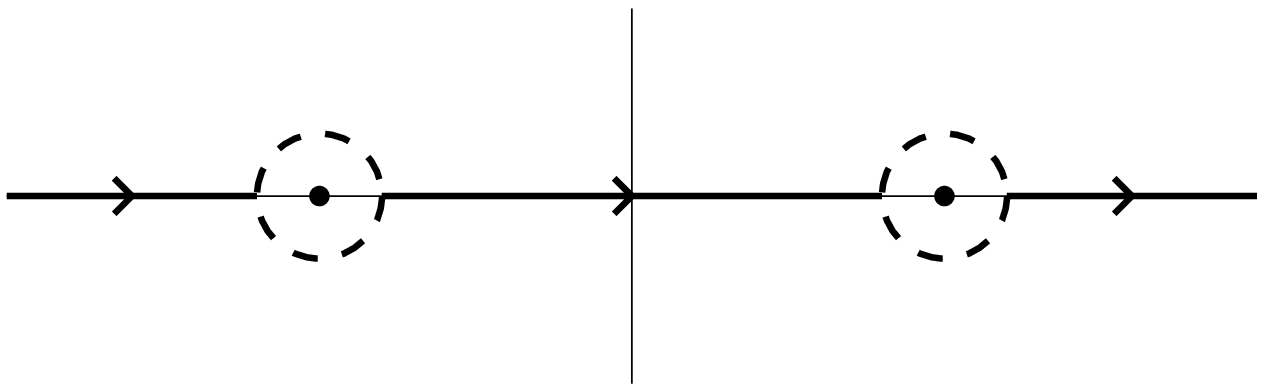}
\end{centre}
We can go either above or below each singularity.  This gives us four
different propagators.  They all have different properties.

If we go above each singularity then it is clear that the integral
vanishes for $x_0>0$ because we can complete the contour by a
semi-circle in the upper-half plane and this gives $0$ by Cauchy's
theorem.  Thus the propagator vanishes for all $x_0>0$.  We assumed
our propagators were invariant under the connected Lorentz group so
the propagator actually vanishes outside the lower lightcone.  Hence,
this is called the {\bf advanced propagator} and is denoted by
$\Delta_A(x)$.\footnote{Actually, in even dimensions larger than $2$
the advanced propagator also vanishes inside the forward lightcone.
This can be proved by clapping your hands and noticing the sound is
sharp.  This isn't true in odd dimensions which can be shown by
dropping stones into ponds and noticing there are lots of ripples.}

If we go below each singularity then, as above, we can show that the
propagator vanishes outside the upper lightcone.  This is therefore
called the {\bf retarded propagator} and is denoted by $\Delta_R(x)$.
The advanced and retarded propagators are related by
$$\Delta_A(x) = \Delta_R(-x)$$
If we chose to go below the first singularity and above the second
singularity then we will end up with {\bf Feynman's propagator} which is
sometimes denoted by $\Delta_F(x)$ or simply $\Delta(x)$.  This
satisfies
$$\Delta_F(x) = \Delta_F(-x)$$
This propagator looks more complicated than the previous two because
it doesn't have nice vanishing properties.  However, it is the correct
choice of propagator for QFT because its singularities are
sufficiently tame to allow distribution products like
$\F(\Gamma)\Delta(x_i-x_j)$ to make sense.

If we chose to go above the first singularity and below the second we
get {\bf Dyson's propagator}.  In most ways this is similar to
Feynman's propagator.

All four propagators are related by the formula
$$\Delta_A + \Delta_R = \Delta_F + \Delta_D$$
This is easy to show by looking at the contours defining each
distribution.

Finally, we should explain how to form the propagators in a massless
theory.  To do this we consider the quadratic form on spacetime as a
map $\RRR^{1,n-1} \longrightarrow \RRR$.  We then take the massless
propagator on $\RRR$ and pullback to a distribution on $\RRR^{1,n-1}-0$
using the quadratic form.  This is a homogeneous distribution on
$\RRR^{1,n-1}-0$ and so using previous results we can extend this to a
homogeneous distribution on $\RRR^{1,n-1}$ if $n>2$.  We get the same
problem as before for dimensions $1$ and $2$ --- the extension may not
be unique or homogeneous.

\subsection{Wavefronts and Distribution Products}

In general it is not possible to define the restriction of a
distribution to some submanifold of its domain of definition nor is it
possible to define the product of two distributions if they have
singularities in common.  The theory of wavefront sets allows us to
give a sufficient condition when these two operations can be
performed.  This is important for us because we need to take the
product of distributions in our definition of a renormalisation
prescription.

Define two distributions by
$$f_1(\varphi) = \int_{\C_1} {\varphi(x) \over x} dx \qquad
f_2(\varphi) = \int_{\C_2} {\varphi(x) \over x} dx$$
Where the two contours are shown below
\begin{centre}
\includegraphics[scale=.6]{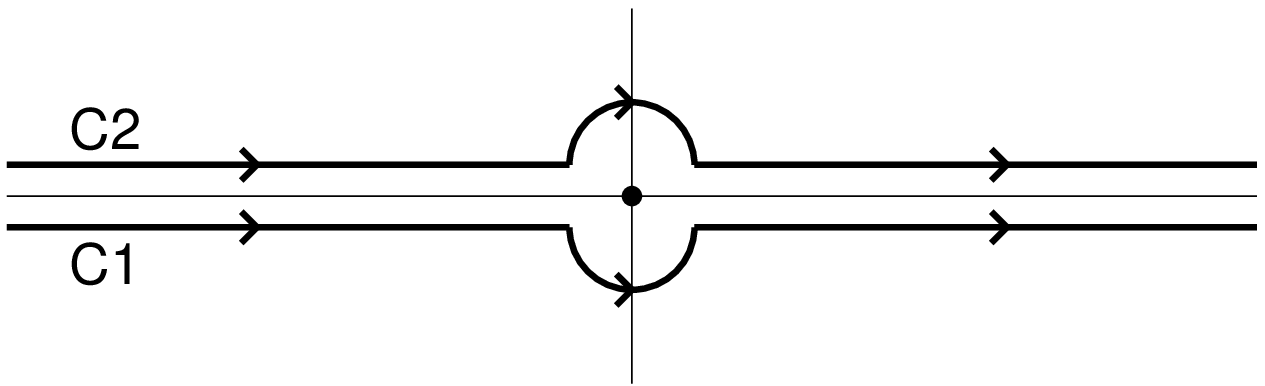}
\end{centre}
If we could define products like $f_1f_1$ or $f_1f_2$ we might expect
that the satisfy equations of the form $\widehat{f_1f_2} = \hat f_1 *
\hat f_2$.  In fact, provided that the convolution of the two
distributions $\hat f_1$ and $\hat f_2$ can be defined we could take
the inverse Fourier transform as the definition of the product of
distributions.  Let us therefore compute the Fourier transforms of
$f_1$ and $f_2$
$$\hat f_1(p) = \begin{cases}1 & \hbox{if $p<0$} \\
0 & \hbox{if $p>0$}\end{cases} \qquad\qquad
\hat f_2(p) = \begin{cases}0 & \hbox{if $p<0$} \\
1 & \hbox{if $p>0$}\end{cases}$$
It is now easy to see that we can compute $\hat f_1 * \hat f_1$ and
$\hat f_2 * \hat f_2$ but not $\hat f_1 * \hat f_2$.  It is easy to
see that the reason for this is due to the different supports of $\hat
f_1$ and $\hat f_2$.  So we should expect that $f_1f_1$ and $f_2f_2$
exist but $f_1f_2$ doesn't.  It is easy to show:

{\it If $f$ and $g$ have support in $x\ge M$ then $f*g$ can be
defined (similarly, if $f$ and $g$ have support in $x\le M$).
However, if $f$ has support in $x\ge M$ and $g$ has support in $x\le
N$ then there could be convergence problems defining $f*g$.}

A little more thought shows that it isn't where the support is that is
the problem but where the Fourier transform is not rapidly
decreasing.  This idea leads to the concept of a {\bf wavefront set}.

Suppose that $u$ is a distribution on $\RRR^n$.  Define $\Sigma(u)$ to
be the cone (excluding $0$) in $\RRR^{n*}$ of directions in which
$\hat u$ does not decrease rapidly.  So, $\Sigma(u)$ measures the
global smoothness of $u$.  However, we know that the only problem with
distribution products is when singularities occur at the same point,
so we need some kind of local version of $\Sigma(u)$.

Let $\psi$ be a bump function about the point $x$.  Then
$$\Sigma_x(u) = \bigcap_\psi \Sigma(u\psi)$$
is a cone which depends only on the behaviour of $u$ near $x$.  The
wavefront set is defined to be pairs of points $(x,p)\in \RRR^n \times
\RRR^{n*}$ such that $p \in \Sigma_x(u)$.  This set is denoted by
$\wf(u)$.\footnote{Note that the wavefront set is a subset of the
cotangent bundle of $\RRR^n$}

For example, it is easy to find the wavefront sets for $f_1$ and $f_2$
$$\wf(f_1) = \{ (0,p) : p<0 \} \qquad\qquad \wf(f_2) = \{ (0,p) : p>0 \}$$

The main result on wavefront sets that we use is the following one
about when it is possible to pullback distributions

{\it Suppose $f:M\longrightarrow N$ is a smooth map between smooth
manifolds and $u$ is a distribution on $N$.  Suppose there are no points
$(x,p)\in \wf(u)$ such that $p$ is normal to $df(T_xM)$.  Then the
pullback $f^*u$ can be defined such that $\wf(f^*u) \subset f^*\wf(u)$.
This pullback is the unique, continuous way to extend pullback of
functions subject to the wavefront condition.}

We shall briefly explain how to pullback a distribution.  Any
distribution can be given as a limit of distributions defined by
smooth compactly supported functions.  Pick such a sequence $u_j$ for
the distribution $u$.  We can further assume that this sequence
behaves uniformly well in directions not contained in the wavefront.
Having done this we can pullback the $u_j$ to functions on $M$.  The
condition on the normal directions missing the wavefront is then
enough to guarantee that the pullback functions converge to a
distribution.

This result allows us to define the product of distributions in the
case when their wavefront sets have certain properties.  If we have
two distributions $u$ and $v$ on $M$ then we can form the
distribution $u\tensor v$ on $M\times M$ in the obvious way.  If we
pullback along the diagonal map then this will be the product of the
two distributions.  The restriction can be done provided that there
are no points $(x,p)\in \wf(u)$ such that $(x,-p)\in \wf(v)$.

We can also use the above theorem to restrict distributions to
submanifolds

{\it If $M\subset N$ is a smooth submanifold of $N$ and $u$ is a
distribution on $N$ then it is possible to restrict $u$ to $M$
provided that there are no points $(x,p)\in \wf(u)$ such that $x\in M$
and $p$ tangent to $M$.}

There is a special case of the above result that is sufficient for our
purposes.  Suppose that we specify a proper cone\footnote{This is a
cone which is contained in some half space} $C_x$ at each point $x\in
M$.  Suppose that $u$ and $v$ have the property that if $(x,p)\in \wf$
then $p\in C_x$.  In this case we can multiply the distributions $u$
and $v$.  This clearly follows from the above general result but it
can be shown in an elementary way.

We can assume that $\hat u$ and $\hat v$ are rapidly decreasing
outside some proper cone $C$.  If we picture taking the convolution we
see that there is only a bounded region where the distribution is not
of rapid decay.  Therefore the convolution is well defined and so is
its inverse Fourier transform.  This gives us the product of
distributions.

We can now understand why the Feynman propagator is the correct choice
of propagator for QFT.  If we work out its wavefront set
$\wf(\Delta_F)$ we find that the singularities occur only on the
lightcone.  At the origin, the wavefront is $\RRR^n-0$; at each point
on the forward lightcone the wavefront is the forward lightcone; and at
each point on the backward light cone the wavefront is the backward
lightcone.  This means that $\Delta_F$ satisfies the properties that
allowed it to be multiplied by itself (except at the origin).  The
Dyson propagator is similar except on the forward lightcone the
wavefront is the backward lightcone and on the backward lightcone the
wavefront is the forward lightcone.

The advanced propagator has wavefront supported only on the forward
lightcone.  At each point on the forward lightcone the wavefront is a
double cone.  The retarded propagator is similar except it is
supported on the backward lightcone.

These wavefronts show that the propagators we have studied are special
because their wavefront sets are about half the expected size.  There
are only really two other possibilities: the wavefront cound be
supported on the full lightcone and be the forward lightcone at each
point or it could be supported on the full lightcone and be the
backward lightcone at each point.  These turn out to be ordinary (not
fundamental) solutions to the differential equation.  They correspond
to the Fourier transforms of the delta functions $\delta_{S_1}$ and
$\delta_{S_2}$.

\eject\section{Higher Dimensional QFT}

We have now seen in detail the combinatorics that occur in $0$
dimensional QFT and how to solve many of the divergence problems.  All
these problems will still exist in higher dimensional space-time,
however there are new problems that could not be seen in the $0$
dimensional case.

Recall that we attached a space--time position $x_i$ to each vertex
and a propagator $\Delta(x_i-x_j)$ to each edge in the Feynman
diagram.  These didn't appear in the $0$ dimensional case because
there was only one point in space-time and the propagator was just the
constant function $1$.  We then had to integrate this attached
function over all the possible values of the space-time position for
the valence $4$ vertices (again this doesn't show up in $0$ dimensions
because the integral is trivial).  The problem with these integrals is
that they are usually very divergent.  In this section we will see how
to deal with these new divergences.

\subsection{An Example}

In order to illustrate what we are going to do to regularise the
integrals we will demonstrate the ideas on the following integral
$$\int_\RRR {1 \over |x|} t(x) dx, \qquad\hbox{$t(x)$ smooth with
compact support}$$ If $t(x)$ is supported away from $0$ then this is
well defined and so the integral gives a linear function from such
$t$'s to the real numbers.  We wish to find a linear functional on the
smooth functions with compact support which agrees with the above
integral whenever it is defined.  Recall that linear functionals on
the space of smooth functions with compact support are called
distributions.  Note that,
$${1 \over |x|} = {d \over dx} \(\sign(x)\log|x|\)$$
So, by integration by parts
$$\int_\RRR {1 \over |x|} t(x) dx = -\int_\RRR \sign(x)\log|x| t'(x)
dx$$
The integral on the right hand side is well defined for any
$t(x)$ smooth with compact support; it is the definition of the
distribution
$${d \over dx} M_{\sign(x)\log|x|}$$
So, we could define our original integral to be this distribution.
Before doing this we should think about whether this extension is unique.
Unfortunately, the answer is clearly that it isn't --- we can add any
distribution supported at $0$.  We therefore have a whole family of possible
lifts, perhaps there is a cannonical way to choose one member of the
family?

To try to pick a cannonical lift we should look for various properties
that $1/|x|$ has and try to ensure that the distribution also
has these properties.  One property is that $1/|x|$
is homogeneous of degree $-1$.  Let the {\bf Euler operator} be
given by $E = x {d \over dx}$ and denote by $f$ the distribution we
defined above.  It is then easy to check that
$$(E+1)f = 2\delta(x) \qquad\hbox{and}\qquad (E+1)^2 f = 0$$

As the $n$--th derivative of the delta function is homogeneous of
degree $-1-n$ it is clear that there is no distribution lifting
$1/|x|$ that is homogeneous of degree $-1$.  We can however
insist that it is {\bf generalised homogeneous of degree $-1$} ---
i.e. that it satisfies $(E+1)^2f = 0$.  This limits the possible
family of distributions to
$$f + k\cdot \delta(x)$$
Is there now a canonical choice for $k$?  We will show that there
isn't by showing that rescaling the real line causes the choice of $k$
to change.  Consider the change of scale $x \rightarrow \lambda x$,
for positive $\lambda$
\begin{eqnarray*}
{d \over dx} M_{\sign(x)\log |x|} &\rightarrow& {d \over d\lambda x}
M_{\sign(\lambda x) \log |\lambda x|} \\
&=& {1 \over \lambda} {d \over dx} M_{\sign(x) \log |\lambda x|} \\
&=& {1\over \lambda} {d \over dx} M_{\sign(x) \log |x|} +
{\log|\lambda| \over \lambda} \delta(x)
\end{eqnarray*}
Hence there is no canonical choice for $k$.  However, we have seen
that there is a canonical family of choices which are transitively
permuted by a group of automorphisms.

We could also try to extend the function $1/x$ to a distribution.
This time we would find that it was canonically equal to the
distribution $M'_{\log |x|}$.  The reason for the difference in the
two cases is that $1/|x|$ is homogeneous of degree $-1$ and even (just
like $\delta(x)$) whereas $1/x$ is homogeneous of degree $-1$ and
odd.  In the case where there is a distribution supported at the
origin with the same transformation properties as the function it is
impossible to canonically tell them apart and so there will be a
family of possible lifts.

\subsection{Renormalisation Prescriptions}

We saw in the example that one way to remove singularities that occur
in the integrals is to regard them as distributions rather than
functions.  The problem with this was that there was not necessarily a
canonical lift, however we can hope that there is a canonical family
of lifts which are permuted transitively by symmetries.

Let us suppose we have some map $\F$ which associates to a Feynman
diagram $\Gamma$ a distribution $\F(\Gamma)$ with the following
properties
\begin{enumerate}\parskip 0pt
\item The distribution $\F(\Gamma)$ lifts the naive function 
$\prod\limits_{i\leftrightarrow j} \Delta(x_i,x_j)$, associated to the Feynman
graph $\Gamma$, whenever this function is defined
\item $\F(\bullet) = 1$
\item If $\sigma(\Gamma_1) = \Gamma_2$ is an isomorphism then
$\sigma(\F(\Gamma_1)) = \F(\Gamma_2)$
\item $\F(\Gamma_1 \sqcup \Gamma_2) = \F(\Gamma_1)\F(\Gamma_2)$
\item $\F(\Gamma)$ should be unchanged if we replace all $x_i$ with
$x_i+x$
\item Let $\Gamma \cup e$ be a graph with a marked edge joining
vertices $i$ and $j$.  Then, if $x_i\ne x_j$, $\F(\Gamma\cup e) =
\Delta(x_i,x_j) \F(\Gamma)$ whenever this is well defined
\end{enumerate}
Suppose we have some map $\F$ with these properties and we know its
value for all graphs $\Gamma'$ contained within $\Gamma$.  What can we
say about $\F(\Gamma)$?
\begin{itemise}\parskip 0pt
\item If $\Gamma$ is not connected then $\F(\Gamma)$ is determined as
the product $\F(\Gamma_1)\F(\Gamma_2)$, where $\Gamma = \Gamma_1
\sqcup \Gamma_2$.
\vskip 7pt
Note that this is independent of the decomposition of $\Gamma$ into
components.
\item Suppose $\Gamma$ is connected and there is some $x_i\ne x_j$,
then we can find $k$,$l$ with $x_k\ne x_l$ and $k$,$l$ joined by an
edge in $\Gamma$.  Therefore, $\F(\Gamma) = \F(\Gamma-e)\Delta(x_k,x_l)$.
\vskip 7pt
Note this is independent of the choice of edge we make.  However, we
have already seen that if $\Delta(x)$ has singularities outside $x=0$
this product may not make sense.  We will later show by induction that
choosing the Feynman propagator gives sufficiently nice wavefront sets
to avoid this problem.
\end{itemise}
Thus, $\F(\Gamma)$ is determined except on the diagonal $x_1=x_2 =
\cdots$.  By translation invariance we can therefore regard the
distribution $\F(\Gamma)$ as living on some space minus the origin.
We saw in the previous section that it is quite often possible to
extend such distributions to distributions on the whole space.  The
distributions which occur in Physics will all be extendible in this
way.\footnote{If we worked with
hyperfunctions instead of distributions then we could always do this
extension.}  Of course, the definition of $\F(\Gamma)$ is not unique
as it can be changed by any polynomial in the derivatives of the $x_i$
acting on a distribution supported on the diagonal.  This is exactly
the same non-uniqueness we found when trying to extend $1/|x|$.

So, we have seen that the axioms for a renormalisation prescription
make sense, although they certainly do not uniquely define the value
of $\F(\Gamma)$.  This makes the above inductive definition of
$\F(\Gamma)$ useless for doing real calculations; for these we want
some more algorithmic way to determine the diagonal term (these can be
found in Physics books under names like dimensional regularisation,
minimal subtraction, Pauli--Villers regularisation...).  

\subsection{Finite Renormalisations}

Before proceeding we want to change, slightly, the definition of
Feynman diagram.

We label the ends of edges with fields or derivatives of fields.
Recall that vertices in Feynman diagrams came from terms in the
Lagrangian --- in our example of $\varphi^4$, this term gave rise to
valence $4$ vertices.  In more complicated Lagrangians we may have
terms like $\varphi^2\d\varphi$ which should give rise to a valence
$3$ vertices with edges corresponding to $\varphi$, $\varphi$ and
$\d\varphi$\footnote{These Feynman diagrams still look different from
the ones in the physics literature however the relationship is easy.
Physicists use different line styles to indicate the edge labels.
They also implicitly sum over many different graphs at once --
e.g. summing over all possible polarisations of a photon.  So a
physics Feynman diagram corresponds usually to a sum of diagrams of
our form}.

We need to decide what propagator to assign to an edge with ends
labeled by $\D_1 \varphi_1$ and $\D_2 \varphi_2$ (where $\D$ are
differential operators).  The answer is quite simple,
$$\D_1\D_2 \Delta_{\varphi_1,\varphi_2}(x_1,x_2)$$
where $\Delta_{\varphi_1,\varphi_2}$ is some basic propagator that can
be computed in a similar way to how we first obtained $\Delta$.

Let $M$ be the vector space which is space-time.  Recall that, to a
Feynman diagram $\Gamma$ with $|\Gamma|$ vertices we attach a
distribution $\F(\Gamma)$ on the space $M^{\oplus |\Gamma|}$, which
we denote by $M^\Gamma$.  By $\lieU(M)$ we mean the universal
enveloping algebra of the Lie algebra generated by ${\d \over \d
x_i}$.  By $\lieU(M)^\Gamma$ we mean the tensor product of $|\Gamma|$
copies of $\lieU(M)$.

A {\bf finite renormalisation} is a map $\C$ from Feynman diagrams to
polynomials in $\d \over \d x_i$ which is
\begin{enumerate}\parskip 0pt
\item Linear in the fields labeling the edges
\item $\C(\bullet) = 1$
\item $\C$ commutes with graph isomorphisms
\item $\C(\Gamma_1 \sqcup \Gamma_2) = \C(\Gamma_1)\C(\Gamma_2)$
\end{enumerate}
Physicists call $\C$ the {\bf counter term}.  Note that $\C(\Gamma)$ is
an element of $\lieU(M)^\Gamma$.

We can make a finite renormalisation act on a renormalisation
prescription as follows
$$\C[\F](\Gamma) = \sum_{\gamma\subset \Gamma} \C(\gamma)
\F(\C(\gamma)^{-1} \Gamma/\gamma)$$
{\bf Warning.}  There is a lot of implicit notation in the above
definition which is explained below.

$\gamma\subset \Gamma$ is required to contain all vertices of
$\Gamma$, although it need not contain all edges.  $\Gamma/\gamma$ is
the graph obtained from $\Gamma$ by contracting the components of
$\gamma$ to single points (note that the edges in $\gamma$ contract
too).

$\C(\gamma)$ is a polynomial in ${\d \over \d x_i}$; as explained
 above these are elements of a universal enveloping algebra
(in this case the Lie algebra is Abelian although it doesn't
cause problems to consider non-Abelian Lie algebras).  Universal
enveloping algebras are naturally Hopf algebras --- i.e. they have
three more operations: counit ($\eta$), antipode ($S$) and
comultiplication ($\Delta$).  In our case
$$\Delta\({\d \over \d x_i}\) = {\d \over \d x_i} \tensor 1 + 1 \tensor
{\d \over \d x_i}$$
and
$$S\({\d \over \d x_i}\) = - {\d \over \d x_i}$$
For an element $g$ of a Hopf algebra the symbol $g^{-1}$ is shorthand
for $S(g)$, as the antipode is a kind of linearised inverse.  If an
element $g$ occurs twice in an expression then there is an implicit
summation convention.  Suppose that
$$\Delta(g) = \sum_i g_{1,i} \tensor g_{2,i}$$
then an expression of the form
$$g\hbox{---stuff---} g\hbox{---stuff---}$$
is shorthand for
$$\sum_i g_{1,i}\hbox{---stuff---} g_{2,i}\hbox{---stuff---}$$
If the element $g$ occurs $3$ times in an expression then we
appear to have two choices for the comultiplication to use: $(\id
\tensor \Delta)\circ\Delta$ or $(\Delta \tensor \id)\circ\Delta$.
Fortunately, the axiom of co-associativity gives that these give the
same result.  Thus, we denote by $\Delta^2$ either of these
operations.  Similarly, we get operations
$$\Delta^n : G \longrightarrow G^{\tensor (n+1)}$$
Just as above, we use these operations to give meaning to an
expression with $g$ occuring $n+1$ times.

We now need to explain how an element of $\lieU(M)^\Gamma$ acts on a
Feynman graph $\Gamma$.  We can think of elements of $\lieU(M)^\Gamma$
as being differential operators attached to the vertices of the
Feynman graph $\Gamma$.  We let a differential operator attached to
the vertex $v$ act in the following Leibnitz--like way:

$$
\begin{minipage}{15mm}
\begin{fmfgraph*}(10,10)
\fmfleft{i}
\fmfrightn{o}{2}
\fmf{plain}{i,v}
\fmf{plain}{v,o1}
\fmf{plain}{v,o2}
\fmflabel{$a$}{i}
\fmflabel{$b$}{o1}
\fmflabel{$c$}{o2}
\fmfdot{v}
\end{fmfgraph*}
\end{minipage}
\longrightarrow\ \ \ \ \ \ \ 
\begin{minipage}[c]{15mm}
\begin{fmfgraph*}(10,10)
\fmfleft{i}
\fmfrightn{o}{2}
\fmf{plain}{i,v}
\fmf{plain}{v,o1}
\fmf{plain}{v,o2}
\fmflabel{$\d a$}{i}
\fmflabel{$b$}{o1}
\fmflabel{$c$}{o2}
\fmfdot{v}
\end{fmfgraph*}
\end{minipage}
+\ \ \ \ \ \ 
\parbox{15mm}{\begin{fmfgraph*}(10,10)
\fmfleft{i}
\fmfrightn{o}{2}
\fmf{plain}{i,v}
\fmf{plain}{v,o1}
\fmf{plain}{v,o2}
\fmflabel{$a$}{i}
\fmflabel{$\d b$}{o1}
\fmflabel{$c$}{o2}
\fmfdot{v}
\end{fmfgraph*}}
+\ \ \ \ \ \ 
\parbox{15mm}{\begin{fmfgraph*}(10,10)
\fmfleft{i}
\fmfrightn{o}{2}
\fmf{plain}{i,v}
\fmf{plain}{v,o1}
\fmf{plain}{v,o2}
\fmflabel{$a$}{i}
\fmflabel{$b$}{o1}
\fmflabel{$\d c$}{o2}
\fmfdot{v}
\end{fmfgraph*}}
$$
\vskip 3mm
So, we end up with a sum of Feynman graphs rather
than a single Feynman graph.  We extend the action to sums of Feynman
graphs by linearity (and similarly for $\F$).  When applying the
differential operators in the definition of the action
there is an additional convention: Any edge which occurs in $\gamma$
is ignored when applying the differential operator to $\Gamma$.

$\F(\C(\gamma)^{-1} \Gamma/\gamma)$ is a distribution on
$M^{\Gamma/\gamma}$ rather than on $M^\Gamma$.  We, need
it to be a distribution on the bigger space.  This is easy to do
because $M^{\Gamma/\gamma}$ is a closed subspace of $M^\Gamma$.
Suppose that $D$ is a distribution on a closed subspace $X$ of $Y$
then we can extend $D$ to a distribution $D'$ on $Y$ by
$$D'(f) := D(f|_X)$$

Finally, we need to say how $\C(\gamma)$ acts on distributions.  This
is just the usual action of differentiation on distributions.

We need to check that $\C[\F]$ is indeed a renormalisation
prescription.  Most of the axioms are trivially satisfied
\begin{itemise}\parskip 0pt
\item $\C[\F]$ commutes with graph isomorphisms, obviously
\item $\C[\F]$ is multiplicative on disjoint unions of graphs,
obviously
\item $\C[\F]$ is translation invariant, obviously
\item $\C[\F](\Gamma \cup e) = \C[\F](\Gamma) \C[\F](e)$ for $e$ an
edge from $i$ to $j$ and $x_i\ne x_j$.  This final axiom requires a
little more work than the previous ones.  As we care only about $x_i
\ne x_j$ we can ignore terms in the sum for which $e$ is an edge in
$\gamma$.  Hence
\begin{eqnarray*}
\C[\F](\Gamma \cup e) &=& (\hbox{ignored}) + \sum_{\gamma \subset
\Gamma} \C(\gamma) (\F(\C(\gamma)^{-1}(\Gamma \cup e)/\gamma) \\
&=& (\hbox{ignored}) + \sum_{\gamma \subset\Gamma} \C(\gamma)
\F(\C(\gamma)^{-1}\Gamma/\gamma) \F(\C(\gamma)^{-1}e)) \\
&=& (\hbox{ignored}) + \sum_{\gamma \subset\Gamma} \C(\gamma)
\F(\C(\gamma)^{-1}\Gamma/\gamma)\times \C(\gamma)\C(\gamma)^{-1} \F(e)
\\
&=& (\hbox{ignored}) + \sum_{\gamma \subset\Gamma} \C(\gamma)
\F(\C(\gamma)^{-1}\Gamma/\gamma)\times \eta(\C(\gamma)) \F(e) \\
&=& (\hbox{ignored}) + \sum_{\gamma \subset\Gamma} \C(\gamma)
\F(\C(\gamma)^{-1}\Gamma/\gamma)\times \F(e) \\
&=& (\hbox{ignored}) + \C[\F](\Gamma)\C[\F](e)
\end{eqnarray*}
\end{itemise}

\subsection{A Group Structure on Finite Renormalisations}

If we just compute the composition of two finite renormalisations on a
renormalisation prescription we get
\begin{eqnarray*}
\C_1[\C_2[\F]](\Gamma) &=& \sum_{\gamma_1 \subset \Gamma}
\C_1(\gamma_1) \C_2[\F](\C_1(\gamma_1)^{-1} \Gamma/\gamma_1) \\
&=& \sum_{\gamma_1 \subset \Gamma} \sum_{\gamma_2 \subset
\C_1(\gamma_1)^{-1} \Gamma/\gamma_1} \C_1(\gamma_1)\C_2(\gamma_2)
\F(\C_2(\gamma_2)^{-1} \C_1(\gamma_1)^{-1} \Gamma/\gamma_1/\gamma_2)
\end{eqnarray*}
This last sum can be written in the form
$$\sum_{\gamma\subset \Gamma} \C(\gamma) \F(\C(\gamma)^{-1}
\Gamma/\gamma)$$
where $\C$ is defined by
$$\C(\Gamma) = \sum_{\gamma\subset \Gamma} \C_1(\gamma)
\C_2(\C_1(\gamma)^{-1} \Gamma/\gamma)$$
This is true by a simple calculation:
\begin{eqnarray*}
\sum_{\gamma\subset \Gamma} \C(\gamma)
\F(\C(\gamma)^{-1}\Gamma/\gamma)\!\! &=& \sum_{\gamma \subset \Gamma}
\sum_{\lambda \subset \gamma} \C_1(\lambda) \C_2(\C_1(\lambda)^{-1}
\gamma/\lambda) \F(\C_2(\C_1(\lambda)^{-1} \gamma/\lambda)^{-1}
\C_1(\lambda)^{-1} \Gamma/\gamma) \\
&=& \sum_{\lambda \subset \Gamma} \sum_{\lambda \subset \gamma \subset
\Gamma} \C_1(\lambda) \C_2(\C_1(\lambda)^{-1}
\gamma/\lambda) \F(\C_2(\C_1(\lambda)^{-1} \gamma/\lambda)^{-1}
\C_1(\lambda)^{-1} \Gamma/\gamma) \\
&=& \sum_{\lambda \subset \Gamma} \sum_{\gamma' \subset \Gamma/\lambda}
\C_1(\lambda) \C_2(\C_1(\lambda)^{-1}
\gamma') \F(\C_2(\C_1(\lambda)^{-1} \gamma')^{-1}
\C_1(\lambda)^{-1} \Gamma/\lambda/\gamma') \\
&=& \sum_{\lambda \subset \Gamma} \sum_{\gamma' \subset
\C_1(\lambda)^{-1}\Gamma/\lambda}
\C_1(\lambda) \C_2(\gamma') \F(\C_2(\gamma')^{-1}
\C_1(\lambda)^{-1} \Gamma/\lambda/\gamma')
\end{eqnarray*}
This suggests that we could try to define a product on renormalisation
by
$$(\C_1\circ \C_2) := \C$$
We will now show that this defines a group structure on the set of
renormalisations.

{\bf Associativity.}  If renormalisations acted faithfully on
renormalisation prescriptions then this would follow immediately.
Unfortunately, they do not act faithfully.  Consider instead the set
of all maps $\F$ from Feynman diagrams to distributions which commute
with graph isomorphism.  We show that the renormalisations act on this
set faithfully and hence the product is associative.  Define $\F$ by
$$\F(\Gamma) =
\begin{cases} \delta(x) & \hbox{when $\Gamma = \bullet$} \\
0 & \hbox{otherwise}
\end{cases}$$
Then it is easy to see that
$$\C[\F](\Gamma) = \C(\Gamma)\F(\bullet)$$
and so the action is faithful.

{\bf Identity.}  The identity renormalisation is clearly
$$\C(\Gamma) =
\begin{cases} 1 & \hbox{if $\Gamma$ is a union of points} \\
0 & \hbox{otherwise}
\end{cases}$$

{\bf Inverse.}  Given $\C_2$ we want to find a $\C_1$ such that
$$\sum_{\gamma \subset \Gamma} \C_1(\gamma) \C_2(\C_1(\gamma)^{-1}
\Gamma/\gamma)=0$$
whenever $\Gamma$ is not a union of points.  We define $\C_1$ by
induction on graph size as
$$\C_1(\Gamma) = -\sum_{\gamma\varsubsetneq \Gamma} \C_1(\gamma)
\C_2(\C_1(\gamma)^{-1} \Gamma/\gamma)$$
It is easy to see this gives an inverse.

So, the product defined above does indeed give a group structure to
the set of renormalisations and these renormalisations act on the set
of all renormalisation prescriptions.  Importantly, this action turns
out to be transitive.  To show this, suppose that we have two
prescriptions $\F_1$ and $\F_2$ and a connected graph $\Gamma$ such that
$$\F_1(\Gamma) \ne \F_2(\Gamma)\qquad\hbox{but}\qquad \F_1(\gamma) =
\F_2(\gamma)\hbox{ for $\gamma\subset\Gamma$}$$
Define a renormalisation $\C$ on connected graphs $\gamma$ to be
\begin{itemise}\parskip 0pt
\item $1$ if $\gamma$ is a point
\item If $\gamma = \Gamma$, then $\F_2(\Gamma)-\F_1(\Gamma)$ is a
translation invariant distribution supported on the diagonal of
$M^\Gamma$.  This can be written in the form
$\C(\Gamma)\delta(\hbox{diag})$.  Use this to define $\C(\Gamma)$.
\item $0$ if $\gamma$ not a point or $\Gamma$
\end{itemise}
Extend this to all graphs using the multiplicativity under disjoint
union.  This is a renormalisation.  It is easy that $\C[\F_1](\gamma)
= \F_2(\gamma)$ for all $\gamma \varsubsetneq \Gamma$.  Consider now
\begin{eqnarray*}
\C[\F_1](\Gamma) &=& \sum_{\gamma\subset \Gamma} \C(\gamma)
\F_1(\C(\gamma)^{-1} \Gamma/\gamma) \\
&=& \F_1(\Gamma) + \C(\Gamma)\F_1((\C(\Gamma)^{-1}\Gamma)/ \Gamma) \\
&=& \F_1(\Gamma) + \C(\Gamma)\delta(\hbox{diag})
\end{eqnarray*}
The equality between lines $2$ and $3$ is because the only non-zero
term in the implicit sum is when none of the derivatives act on the
graph.  This is exactly the remaining term in line $3$.

Hence, by repeating this construction a finite number of times we find
a renormalisation $\C$ for which $\C[\F_1] = \F_2$ for all connected
graphs $\gamma \subset \Gamma$.  Also, $\C[\F_1] = \F_1$ for all
connected graphs distinct from $\Gamma$.  Hence, $\C[\F_1]$ agrees with
$\F_2$ on a larger number of graphs but only changes from $\F_1$ on a
finite number of connected graphs.  Hence, it is possible to repeat
this construction an infinite number of times and compose all the
$\C$.  This composition is well defined because only a finite number
of the $\C$ act non-trivially on any graph.  Thus we have constructed
a renormalisation $\C$ such that $\C[\F_1] = \F_2$ and so {\it the
group of renormalisations acts transitively on the set of
renormalisation prescriptions}.

\subsection{More Conditions on Renormalisation Prescriptions}

We now add a new condition that we want renormalisation prescriptions
to have.  This means we should go through the proof in the previous
two sections again to check that they still work in the new
framework.  However, this will be left as an exercise for the reader
--- most of the proof only require trivial modifications.

There is an action of $\lieU(M)^\Gamma$ on both Feynman graphs,
distributions on $M^\Gamma$ and $\lieU(M)^\Gamma$ itself.  It is
therefore natural to require that $\C$ and $\F$ both commute with
these actions.  From now on we will assume that this additional axiom
has been imposed on renormalisations and renormalisation
prescriptions.

Given a renormalisation prescription $\F$ what is the subgroup of the
renormalisations that fixes it?  We claim it is the $\C$ which satisfy
either of the following two equivalent conditions
\begin{enumerate}\parskip 0pt
\item $\C(\Gamma)$ acts trivially on the diagonal distribution
$\delta(\hbox{diag})$
\item $\C(\Gamma)$ is in the left ideal generated by the diagonal
embedding of first order differential operators
$$\({\d\over \d x},1,\dots,1\) + \(1,{\d\over \d x},1,\dots,1\) +
\cdots + \(1,\dots,1,{\d\over \d x}\)$$
\end{enumerate}
It is clear that these two conditions are equivalent.

Suppose that $\C$ has property $2$.  Then we can assume that
$\C(\gamma)$ is a sum of terms of the form $ba$, where $b$ is any
differential operators and $a$ is one of the generators of the ideal
defined in property $2$.  Then
\begin{eqnarray*}
\C(\gamma)\F(\C(\gamma)^{-1} \Gamma/\gamma)
&=& \sum ba\F(S(a)S(b)\Gamma/\gamma) \\
&=& \sum baS(a) \F(S(b)\Gamma/\gamma) \\
&=& \sum b \eta(a) \F(S(b)\Gamma/\gamma) \\
&=& 0
\end{eqnarray*}
The equality from line $1$ to $2$ is shown as follows: $S(a)$ is a
differential operator in the ideal and it is easy to check that this
descends to a diagonal differential operator acting on
$\Gamma/\gamma$.  This can be pulled through $\F$ by assumption.
Finally, recall that the distribution on $M^{\Gamma/\gamma}$ is lifted
to one on $M^\Gamma$, it is easy to see that the action of the
diagonal operator lifts to $S(a)$ on the lifted distribution.

By definition $aS(a) = \eta(a)$, and this is zero because $a$ contains
terms which are differential operators.

Hence, the action of $\C$ on $\F$ is given by
$$\C[\F](\Gamma) = \sum_{\gamma\subset \Gamma} \C(\gamma)
\F(\C(\gamma)^{-1} \Gamma/\gamma) = \F(\Gamma)$$
as all terms except when $\gamma$ is a union of points give zero and
the other case gives $\F(\Gamma)$.

Suppose that $\C$ fixes $\F$.  By induction we can assume that $\C$
has either of the above properties for small graphs $\gamma$.  Let
$\Gamma$ be a graph for which we haven't determined the value of
$\C(\Gamma)$ but for which we know $\C(\gamma)$ for all smaller
$\gamma$.  Then
$$\C[\F](\Gamma) = \sum_{\gamma\subset \Gamma} \C(\gamma)
\F(\C(\gamma)^{-1} \Gamma/\gamma) = \F(\Gamma) +
\C(\Gamma)\F(\bullet)$$
Most of the terms are zero by the calculation above.  The last term
shows that $\C(\Gamma)$ must act trivially on $\F(\bullet)$ which is
just the distribution $\delta(\hbox{diag})$.

Thus we have shown that the subgroup of renormalisations which fix any
given renormalisation prescription is as claimed.  Note that this
subgroup is independent of $\F$.  Hence this subgroup is normal and
the quotient of the renormalisations by the stabiliser acts simply
transitively on the renormalisation prescriptions.  Simply transitive
actions of groups on sets are sometimes called {\bf torsors}.

Looking at physics books shows that they require a new restriction on
renormalisation prescriptions: Suppose that $\Gamma\cup e$ is a graph
with an edge $e$ which has ends in different components of $\Gamma$
then $\F(\Gamma\cup e) = \F(\Gamma)\F(e)$ everywhere\footnote{When we
define the wavefront of a graph in the next section it will be easy to
show that this product is well defined}.  The corresponding
restriction we require on $\C$ is that $\C(\Gamma)=0$ if $\Gamma$ is
connected but not 1PI.

\subsection{Distribution Products in Renormalisation Prescriptions}

Recall that the definition of renormalisation prescription required us
to be able to multiply distributions.  For example, if we apply $\F$
to the following graph
$$\parbox{15mm}{\begin{fmfgraph*}(10,10)
\fmfleft{i}
\fmfright{o}
\fmf{phantom}{i,v}
\fmf{plain,left=1,tension=0.1}{v,w}
\fmf{plain,right=1,tension=0.1}{v,w}
\fmfdot{v,w}
\fmf{phantom}{w,o}
\fmflabel{$x$}{v}
\fmflabel{$y$}{w}
\end{fmfgraph*}}
$$
We should get the distribution $\Delta(x-y)\Delta(x-y)$ if $x\ne y$.
It is clear from the condition proved in the previous section that we
want the wavefront set for $\Delta$ to be contained in some half space
at every non-zero point.  The is true for the Feynman (and Dyson)
propagator but is not for the others.  So, we shall always pick the
Feynman propagator as the value for $\F(e)$.  We now need to show that
this choice always allows us to compute the distribution product
$\F(\Gamma)\F(e)$.  We will show this by induction.

To an edge $e$ from $x$ to $y$ we associate a distribution in two
variables $\Delta(x,y)$ which is supposed to be equal to
$\Delta_F(x-y)$.  We should show that this is a well defined
distribution and compute its wavefront.

If we consider the map $f:M^2\longrightarrow M$ given by $(x,y)\mapsto
x-y$ then $\Delta(x,y)$ is the pullback of the distribution
$\Delta_F(z)$ by $f$.  As this map is surjective on the tangent spaces
it clearly satisfies the condition for the pullback to exist.  So,
$\Delta(x,y)$ is a well defined distribution on $M^2$.  The wavefront
is contained in $f^*\wf(\Delta_F)$.  If we just compute this we see
that this is itself contained in
$$\{(x,y,p,-p) : p\succ 0\hbox{ if }x\succ y,p\prec 0\hbox{ if }x\prec
y, p\hbox{ arbitrary if }x=y\}$$
In the above formula we have used the notation that $x\succ y$ if $x-y$
is in the convex hull of the forward lightcone.  Similarly, $x\prec y$
if $x-y$ is in the convex hull of the backward lightcone.  Note that
the set we have described is much bigger than the set
$\wf(f^*\Delta_F)$ however it is small enough for our purpose.

For any graph $\Gamma$ with $m$ vertices we define the following
{\bf wavefront}
$$\wf(\Gamma) = \{(x_1,\dots,x_m,p_1,\dots,p_m) : \sum_{x_i\succcurlyeq x}
p_i \succcurlyeq 0\hbox{ for all $x\in M$}\}$$
Note that the wavefront for the distribution $\Delta(x,y)$ is
contained in what we have defined for the wavefront of the edge
$e$ (even at the origin).

We shall show the following two properties of these wavefront sets.
\begin{enumerate}\parskip 0pt
\item $\wf(\F(\Gamma)) \subset \wf(\Gamma)$ for all graphs $\Gamma$
\item The sets $\wf(\Gamma)$ are nice enough to show that
the distribution products occurring in the definition of a
renormalisation prescription exist
\end{enumerate}
To do this we need the following result on how multiplication of
distributions changes the wavefront
$$\wf(fg) \subset \{(x,p+q): (x,p)\in \wf(f)\hbox{ or }p=0, (x,q)\in
\wf(g)\hbox{ or }q=0\}$$
If $\Gamma=\Gamma_1 \sqcup \Gamma_2$ then it is clear that the product
of distributions $\F(\Gamma_1)\F(\Gamma_2)$ is defined (as they have
no common singularities).  It is also clear that $\wf(\F(\Gamma))
\subset \wf(\Gamma)$.

Suppose $\Gamma\cup e$ is a graph with $e$ an edge from vertex $1$ to
$2$ (without loss of generality).  By convexity of the forward
lightcone it is easy to see that $\wf(\F(\Gamma\cup e)) \subset
\wf(\Gamma\cup e)$.  What we need to check is that the product of
distributions $\F(\Gamma)\F(e)$ is well defined for $x_1\ne x_2$.
Suppose we have two wavefront elements at the same point whose
frequencies sum to zero:
$$(x_1,x_2,\dots,x_m,p_1,p_2,0,\dots,0) \in \wf(\Gamma)$$
and
$$(x_1,x_2,\dots,x_m,-p_1,-p_2,0,\dots,0) \in \wf(e)$$
By picking $x=x_1$ and $x=x_2$ we see that $p_1=p_2=0$ and hence the
two wavefront sets satisfy the sufficient condition to allow the
product $\F(\Gamma)\F(e)$ to exist if $x_1\ne x_2$.

Hence, by induction on the graph size we have shown that the definition
of a renormalisation prescription is well defined.

\eject\section{Renormalisation of Lagrangians}

Recall that we had the following integrals occurring in our calculation
of the Feynman path integral
$$\int J(x_1)\cdots J(x_N) \times \[\int
e^{i\int L_{\rm free}d^nx} \varphi(x_1)^{n_1}\cdots\varphi(x_N)^{n_N}
\D\varphi\] d^nx_1\cdots d^nx_N$$
The inner integral is called a {\bf Green's function} (this is a
slight misnomer because it is a distribution not a function).  If we
can evaluate all the Green's functions then we will basically be done.
To evaluate the Green's function we take a graph with $n_1$ vertices
labeled by $x_1$, \dots, $n_N$ vertices labeled by $x_N$.  From this
graph we will extract (using a chosen renormalisation prescription
$\F$) a distribution.  Summing over all such graphs we will get a
distribution which is the Green's function.

There is an arbitrary choice for the renormalisation prescription $\F$
in the above procedure.  We already know that renormalisation
prescriptions are not unique so this could mean that we end up with
different answers depending on the choice of $\F$.  The one thing we
know about $\F$ is that it is acted on transitively by finite
renormalisations.  This allows us to resolve the apparent
non-uniqueness as follows: Find an action of finite renormalisations
on Lagrangians such that acting simultaneously on the Lagrangian and
$\F$ gives the same Green's functions.  Then, if two physicists pick
different renormalisation techniques they should also pick different
Lagrangians and doing this correctly will allow both physicists to get
the same answers.

In this section we will define actions of finite renormalisations on
both elements of $\V$ and certain generalised Feynman diagrams.

\subsection{Distribution Algebras}

We already have an action of finite renormalisations on distributions
so if we understand the sort of structure that these distributions
posess then it should help us to define actions of finite
renormalisations on these other sets.  In this section we will list
the important structures that these distributions satisfy --- we call
things obeying the resulting axioms {\bf distribution algebras}.

For any finite set $I$ denote by $M^I$ the product $M^{\times |I|}$
with coordinates labeled by $x_i$ for $i\in I$.  Denote by $\dist(I)$
the space of distributions on $M^I$.

There are two obvious operations with finite sets --- disjoint union
and homomorphisms --- we should look at how the spaces of
distributions behave under each of these operations.

If we have distributions $u\in \dist(I)$ and $v \in \dist(J)$ we can get
a distribution in $\dist(I\sqcup J)$ by taking the tensor product
distribution $u\tensor v$.  This gives us maps
$$\dist(I) \times \dist(J) \longrightarrow \dist(I\sqcup J)$$

Consider a map $f:I\rightarrow J$.  What does this do to
distributions?  We can regard $M^I$ as the space of maps from $I$ to
$M$ and similarly for $M^J$; this makes it obvious that the map $f$
induces a map $f:M^J \rightarrow M^I$.  Now, smooth compactly
supported functions on $M^I$ are maps from $M^I$ to $\RRR$.  If $f$ is
not onto we do not get maps $C_0^\infty(M^I) \rightarrow
C_0^\infty(M^J)$ as it is possible to lose the compactness.  There are
several ways around this --- we could restrict to only using onto maps
between sets or we could change the types of distributions we use.
For now we will choose different distributions.  There are two obvious
choices: the compactly supported distributions (the dual space of
$C^\infty$ functions) or the rapidly decreasing distributions (the
dual space of $C^\infty_{\rm poly}$, the at most polynomial increase
functions).  Thus we get a map $F:C^\infty(M^I) \rightarrow
C^\infty(M^J)$ (similarly for $C^\infty_{\rm poly}$).  Finally,
distributions are the dual spaces of these function spaces and so we
get a map $f^*:\dist(J) \rightarrow \dist(I)$.

There is one final property of distributions that we have --- they can
be differentiated.  This amounts to the same thing as saying that
there is an action of $U^I$ on $\dist(I)$ where $U$ is the universal
enveloping algebra of spacetime.  Each map $f:I\rightarrow J$ induces
a map $f^*:U^J \rightarrow U^I$ as follows:  $f$ gives rise to a map
$M^J \rightarrow M^I$; taking the Lie algebra of each side and then
taking the universal enveloping algebra gives rise to the map $f^*$.
This action can also be thought of as arising from the co-product
structure on the spaces $U^I$.

These three structures on the collection of distributions are
compatable in the obvious ways.

So, we define a distribution algebra to be\footnote{In the language of
category theory we have two contravariant tensor functors ${\sf V}:
\underline{\sf Sets} \rightarrow \underline{\sf Vect}$ and ${\sf U}:
\underline{\sf Sets} \rightarrow \underline{\sf Alg}$ and a tensor action
${\sf V}\tensor {\sf U} \rightarrow {\sf V}$}
\begin{itemise}\parskip 0pt
\item A collection of vector spaces, ${\sf V}(I)$, one for each finite set
$I$.
\item For each map $f:I\rightarrow J$ a map $f^*:{\sf V}(J)
\rightarrow {\sf V}(I)$.
\item An action of $U^I$ on ${\sf V}(I)$ which is compatable with $f^*$ in
the sense that
$$f^*(Dv) = f^*(D) f^*(v)$$
where $f^*: U^J \rightarrow U^I$.
\item Natural maps ${\sf V}(I)\times {\sf V}(J) \rightarrow {\sf V}(I\sqcup
J)$ (ie. they are compatable with maps $I\rightarrow I'$ and
$J\rightarrow J'$ and compatable with the action of $U^I$ and $U^J$)
\item Possible extra conditions on the multiplication such as
commutativity, associativity...
\end{itemise}

We can now try to find an action of the finite renormalisations on
$\V$ (possible terms in the Lagrangian).  To do this we will first
define a structure of a distribution algebra on these terms.

Our first guess might be to set ${\sf V}(I) := \V^{\tensor I}$.
Unfortunately this doesn't give rise to induced maps $f^*:{\sf V}(J)
\rightarrow {\sf V}(I)$.  There are two ways around this problem --- we
can simply write down the correct definition of the spaces ${\sf V}(I)$ or
think about category theory for a bit.  We'll start with the category
theory approach.

There is an obvious functor from the category of distribution algebras
to the category of distribution algebras forgetting the existence of
$f^*$ axiom.  Taking the left adjoint of this functor will then give a
functor from this second category to the first (sort of taking the
free/induced distribution algebra).  Applying this left adjoint
functor to the choice $\V^{\tensor I}$ will give us a suitable object.
Thinking about this for a bit allows us to formulate the answer to the
first approach
$$\lag(I) := \bigoplus_{f:I\rightarrow K} U^I \tensor_{U^K}
\V^{\tensor K}$$
We regard $U^I$ as a $U^K$--module via the map $f^*$.  It is obvious
now that given a map $g:I\rightarrow J$ we get a map $g^*:
\lag(J) \rightarrow \lag(I)$ --- given a map $f:J\rightarrow K$ we get
a map $I\rightarrow K$ by composition with $g$.

We now have a distribution algebra structure on the elements of $\V$,
so we should be able to find an action of the finite
renormalisations.  Before doing this we will describe a graphical
notation for elements of $\lag(I)$.

Suppose we have an element of $\V^{\tensor K}$.  We will assume that
it is of the form $v_1\tensor\cdots\tensor v_k$.  For each element of
$K$ draw a solid dot ($\bullet$), usually we will not write the
element of $K$ corresponding to the vertex.  The vertex associated
to $i\in K$ is labeled with the element $v_i$ of $\V$.  We also have a
map $f:I\rightarrow K$.  For each element of $I$ we draw a cross
($\times$).  We draw a dotted line from vertex $i\in I$ to $f(i) \in
K$.  Differential operators act on these crossed vertices subject to
the some relations which allow them to be transfered to the solid
vertices.

For example the following is a possible diagram (with labels on the
solid vertices left out to avoid clutter) from $\lag(\{1,2,3\})$
$$\parbox{25mm}{\begin{fmfgraph}(20,20)
\fmfleftn{i}{4}
\fmfright{o}
\fmf{dots}{i2,v,i3}
\fmf{phantom}{v,w}
\fmf{dots}{w,o}
\fmfdot{v,w}
\fmfv{d.sh=cross,d.size=2thick}{i2,i3,o}
\end{fmfgraph}}$$
An example of how derivatives act is illustrated by the following
(where $\d$ is a first order differential operator from $U$)
$$\parbox{25mm}{\begin{fmfgraph*}(20,20)
\fmfleftn{i}{4}
\fmfright{o}
\fmf{dots}{i2,v,i3}
\fmf{phantom}{v,w}
\fmf{dots}{w,o}
\fmfdot{v,w}
\fmfv{d.sh=cross,d.size=2thick}{i2,i3,o}
\fmfv{l=$\d$}{i3}
\end{fmfgraph*}} +\qquad
\parbox{25mm}{\begin{fmfgraph*}(20,20)
\fmfleftn{i}{4}
\fmfright{o}
\fmf{dots}{i2,v,i3}
\fmf{phantom}{v,w}
\fmf{dots}{w,o}
\fmfdot{v,w}
\fmfv{d.sh=cross,d.size=2thick}{i2,i3,o}
\fmfv{l=$\d$}{i2}
\end{fmfgraph*}} =\quad
\parbox{25mm}{\begin{fmfgraph*}(20,20)
\fmfleftn{i}{4}
\fmfright{o}
\fmf{dots}{i2,v,i3}
\fmf{phantom}{v,w}
\fmf{dots}{w,o}
\fmfdot{v,w}
\fmfv{d.sh=cross,d.size=2thick}{i2,i3,o}
\fmfv{l=$\d$,l.a=70}{v}
\end{fmfgraph*}}
$$
The maps $g^*$ are also easy to visualise with these diagrams.  Place
new crosses for the elements of $J$ and draw dotted lines from these
to the crosses of the element of $I$.  Pull back differential
operators on the vertices of $I$ to the vertices of $J$ using $g^*$
(or the co-product) and then remove the vertices of $I$ along with any
derivatives on them.

If we are not using onto maps $g$ then there is an extra condition on
solid vertices which have no crossed vertices leading to them:  the
labels on these vertices should be considered modulo exact terms in
$\V$.  This can be seen by considering how such a derivative would
pullback through the tensor product $U^I \tensor_{U^K}$.

We can similarly define a distribution algebra structure on Feynman
diagrams.  Again the obvious choice of ${\sf V}(I)$ being graphs on
$|I|$ vertices doesn't allow for the maps $g^*$.  Applying the left
adjoint functor trick again gives us the space $\feyn(I)$.  As before
this can be realised graphically as normal labeled Feynman diagrams on
$|K|$ vertices ($\bullet$).  For any map $f:I\rightarrow K$ we place
crossed vertices ($\times$) for each element of $I$ and join them to
the corresponding vertices of $K$.  Differential operators act on the
crossed vertices as before.  For example, the following could
be an element of $\feyn(\{1,2,3,4\})$.
\vskip-8mm$$\parbox{30mm}{\begin{fmfgraph}(25,25)
\fmfleftn{i}{4}
\fmfrightn{o}{4}
\fmf{dots}{i2,v,i3}
\fmf{plain, left=1,tension=0.5}{v,w,v}
\fmf{plain}{w,x}
\fmf{dots}{o3,x,o2}
\fmfdot{v,w,x}
\fmfv{d.sh=cross,d.size=2thick}{i2,i3,o2,o3}
\end{fmfgraph}}$$
A similar restriction on the labeling of the edges through vertices
with no crossed vertices attached to them applys here too: we should
consider graphs modulo ``exact graphs''.  It is, however, less obvious
when a sum of graphs is exact than when a vertex is exact.

The fact that we take some vertices modulo exact terms suggests that
there is some kind of integration going on (the exact terms would
integrate to zero).  This is indeed the way to think about these
diagrams.  The crossed vertices indicate inserting particles at
spacetime positions indicated by the solid vertices.  They then
propogate according to the diagram but we don't care about how they do
it so we should integrate over all spacetime positions that are
independent of the inserted particles (ie. those that do not have
crossed vertices attached to them).  Eventually we will form a sum
over all possible diagrams with certain insertions, this will
correspond to the amplitude for this event to happen.

\subsection{Renormalisation Actions on $\lag$ and $\feyn$}

In this section we will construct an action of a finite
renormalisation $\C$ on the distribution algebra $\lag$.  To do this
we will first define an action of $\C: \V^{\tensor I} \rightarrow
\lag(I)$.  This will give rise to an action of $\C$ on $\lag$ by
applying the left adjoint functor.  Similarly we will define an action
of $\C$ on $\feyn$ by firstly defining it on the normal Feynman
diagrams and then extending it to $\feyn$ by using the left adjoint
functor.  We will find distribution algebra homomorphisms $\lag
\rightarrow \feyn$ and for each renormalisation prescription $\feyn
\rightarrow \dist$.  These will all be defined to make the following
diagram commute
\begin{diagram}[PostScript=dvips]
\lag(I) &\rTo& \feyn(I)\qquad & & \\
\dTo<{\C} & & \dTo<{\C} & \rdTo(2,1)^{\C[\F]}& \quad\dist(I)\\
\lag(I) &\rTo& \feyn(I)\qquad & \ruTo(2,1)_{\F}& 
\end{diagram}
Given a renormalisation prescription $\F$ the composition $\lag(I)
\rightarrow \dist(I)$ is called a {\bf Green's function}.  In a later
section we will define what we mean by Green's function associated to
a Lagrangian and then we will be able to see that the commutativity of
the above diagram allows us to determine how to change Lagrangians so
as to ballance the effects of choosing different renormalisation
prescriptions.

\subsubsection{The Action on $\lag$}

We firstly define the action of $\C$ on elements of $\V^{\tensor I}$.
Suppose that the element of $\V^{\tensor I}$ is of the form
$v_1\tensor \cdots \tensor v_n$.  We would represent this as a
collection of $n$ solid dots ($\bullet$) with the $i^{th}$ one labeled
by the field $v_i$.  Let $\gamma$ be any graph on these $n$ vertices
which can have its edges labeled using degree $1$ fields from the
fields attached at each vertex.

What do we mean by the last sentence?  Suppose we draw an unlabeled
graph $\gamma$ where the vertex $i$ has valence $d_i$.  Using the
co-commutative co-product on $\V$ we can form the term $\Delta^{d_i}
(v_i) \in \V^{\tensor(d_i+1)}$.  Recall that $\V$ is graded by
assuming that $U$ (the universal enveloping algebra of spacetime) acts
with degree $0$ and the basic fields have degree $1$.\footnote{As an
example, $\varphi$ and $\d\varphi$ are both of degree $1$ but
$\varphi^2$ and $\varphi\d\varphi$ are both of degree $2$} Define a
projection by insisting the last $d_i$ coordinates of
$\Delta^{d_i}(v_i)$ are all of degree $1$.  Label vertex $i$ of
$\gamma$ with the first coordinate of $\Delta^{d_i}(v_i)$ and the
$d_i$ edges with the remaining coordinates.

For example, suppose that a vertex is labeled with $\varphi^3$ and has
valence $1$.  It is easy to compute
$$\Delta^1(\varphi^3) = \varphi^3\tensor 1 + 3\varphi^2\tensor\varphi
+ 3\varphi\tensor\varphi^2 + 1\tensor\varphi^3$$
The projection leaves only the term $3\varphi^2\tensor\varphi$ as the
others have degrees $0$,$2$ and $3$ in their last coordinate.  So we
should label the edge with $\varphi$ and the vertex with $3\varphi^2$.

Thus, we can generate a large (but finite) number of graphs from each
element of $\V^{\tensor I}$.  To each of these graphs $\gamma$ we
obtain a differential operator $\C(\gamma)$ where $\C$ is a finite
renormalisation and it acts on $\gamma$ by ignoring the vertex
labels.  Thus we can form the sum
$$v_1\tensor\cdots\tensor v_n \rightarrow \sum_\gamma
\C(\gamma)\tensor \C(\gamma)^{-1} \tensor\gamma$$
where we are using the previously defined summation convention.

We can think of $\C(\gamma)^{-1}$ as consisting of differential
operators attached to each vertex of $I$.  Hence we can act this on
the graph $\gamma$ by letting it act on the vertex labels (ignoring
the edges).  Now, contract the components of $\gamma$ to points and
multiply the corresponding labels on the vertices together.  This
leads to an element of $\V^{\tensor K}$ where $K$ labels the
components of $\gamma$.  There is a natural map $\pi:I \rightarrow K$
which sends each vertex to the component it is in.  We can pull back
the element of $\V^{\tensor K}$ to an element of $\lag(I)$ using the
map $\pi^*$ (graphically we have a crossed vertex for each original
vertex in $I$ and these are joined to the components of $\gamma$ they
are in).  Finally we can let the differential operators in the
remaining $\C(\gamma)$ act as differential operators in $\lag(I)$
(graphically this means that they act on the crossed vertices).

After all these operations it is not clear that the action of $\C$ on
$V^{\tensor I}$ commutes with the action on $U^I$!  To show this
consider applying the first order differential operator $\d$ to a
vertex.  When we now form the graphs $\gamma'$ we have a choic as to
whether to put the derivative on an edge or on the vertex.  Consider
these two cases separately:

{\bf Derivative on vertex.} The value of $\C(\gamma')$ does not see
the vertex labels and so is equal to the original $\C(\gamma)$.  Thus
we get a term of the form
$$\C(\gamma)\tensor \pi^*(\d\C(\gamma)^{-1}\gamma)$$
{\bf Derivative on edge.} The value of $\C(\gamma')$ is equal to
$\d\C(\gamma)$ because of the assumed compatability of $\C$ under
differentiations.  Taking the co-product of this and applying the
antipode gives two terms
$$\d\C(\gamma)\tensor \pi^*(\C(\gamma)^{-1}\gamma) - \C(\gamma)\tensor
\pi^*(\d\C(\gamma)^{-1}\gamma)$$
This last term cancels out the term from the derivative on a vertex.
The remaining term shows that we have commutativity of the action of
$U^I$.

It now remains to extend this definition of $\C$ to all of $\lag(I) =
\bigoplus U^I\tensor_{U^K} \V^{\tensor K}$.  Do this by acting, with the
above definition of $\C$, on the factor $\V^{\tensor K}$ only
(graphically this would be ignoring the crossed vertices and acting on
the element of $\V^{\tensor K}$ and then re-introducing the crossed
vertices).

We should check that this definition is consistent with the action of
differential operators.  We leave this as an exercise (it basically
follows immediately from the fact that $\C$ commutes with $U^I$).

\subsubsection{The Action on $\feyn$}

As before, we firstly define the action of $\C$ on Feynman diagrams
before using the left adjoint functor to extend this definition to all
of $\feyn$.

Given a Feynman graph $\Gamma$ on vertices $I$ we can form the sum
$$\sum_{\gamma \subset \Gamma} \gamma\tensor \Gamma$$
where the sum is over all subgraphs $\gamma$ of $\Gamma$ which consist
of all vertices of $\Gamma$.  Using the finite renormalisation we can
then form
$$\sum_{\gamma\subset\Gamma} \C(\gamma)\tensor
\C(\gamma)^{-1}\Gamma/\gamma$$
Recall that $\C(\gamma)$ acts on a Feynman graph by differentiating
the labels attached to the edges in a Leibnitz like way and we only
act by differentiation on edges of $\Gamma$ that are not in $\gamma$.
We have a natural map $\pi:I \rightarrow K$ where $K$ are the vertices
of the new Feynman diagram and $\pi$ send an element of $I$ to the
element of $K$ that quotienting by $\gamma$ sends it to.  We can
pullback the Feynman diagram with $\pi^*$ to get an element of
$\feyn(I)$.  Finally we let the remaining copy of $\C(\gamma)$ act as
differential operators on $\feyn(I)$.

As before we should check that the above action commutes with the
action of $U^I$ --- this time we leave it as an exercise.  The map
$\C$ defined above can be extended to a map on $\feyn(I)$ exactly as
the map $\C$ in the previous section was.

Recall that the action of a finite renormalisation $\C$ on $\F$ was
given by
$$\C[\F](\Gamma) = \sum_{\gamma\subset\Gamma}\C(\gamma)
\F(\C(\gamma)^{-1} \Gamma/\gamma)$$
We can regard a renormalisation prescription $\F$ as a distribution
algebra map $\feyn\rightarrow \dist$ because $\F$ already acts on
Feynman diagrams giving distributions and hence using the left adjoint
functor we can extend it.  It is easy to check that we have the
following equality
$$\C[\F] = \F\circ\C$$
where the latter $\C$ is considered as an action on $\feyn$ and the
latter $\F$ is considered as a map $\feyn \rightarrow \dist$ (this is
the commutativity of the triangle in the previous ``commutative
diagram'').

The final map that we need to define is $\lag \rightarrow \feyn$.  As
should be expected by now, we define it on $\V^{\tensor I}$ and then
extend to all of $\lag$ using the left adjoint functor.

Suppose we have an element $v_1\tensor \cdots\tensor v_n$ of
$\V^{\tensor I}$.  Suppose also that each $v_i$ is homogeneous of some
degree $d_i$.  Let $\gamma$ be any Feynman diagram on vertices $I$ for
which vertex $i$ has valence $d_i$.  We label the edges coming from
vertex $i$ with degree $1$ fields from $v_i$ as previously explained.
Note that this time we do not have any fields left to label the vertex
as we use them all on the edges.  We define the map to be the sum of
all such Feynman diagrams:
$$v_1\tensor\cdots \tensor v_n \rightarrow \sum_\gamma \gamma$$
For example the element $\varphi^4\tensor \varphi^2$ would map to
$$48\ \parbox{20mm}{\begin{fmfgraph}(15,15)
\fmfleft{i}
\fmfright{o}
\fmf{plain, left=1,tension=0.5}{i,v,i}
\fmf{plain, left=1,tension=0.5}{v,w,v}
\fmf{phantom}{w,x}
\fmf{plain, left=1,tension=0.5}{x,o,x}
\fmfdot{v,x}
\end{fmfgraph}}
+\qquad
48\ \parbox{20mm}{\begin{fmfgraph}(15,15)
\fmfleft{i}
\fmfright{o}
\fmf{plain, left=1,tension=1}{i,v,i}
\fmf{plain, left=0.5,tension=0.5}{v,o,v}
\fmfdot{v,o}
\end{fmfgraph}}$$

Finally we extend this to $\lag$ (graphically this is done by
re-inserting the crossed vertices).

We leave it as an exercise to check that this is a map of distribution
algebras and the square in the ``commutative diagram'' actually
commutes with this choice of horizontal map.

\subsubsection{An Action on Lagrangians}

In this section we will give the definition of the Green's functions
for a quantum field theory and how there is an action of finite
renormalisations on the Lagrangians which exactly compensates for the
action of $\C$ on Green's functions.

It is unfortunate the the action of $\C$ on $\lag$ is not a
homomorphism
$$\C(v_1\tensor v_2) \ne \C(v_1)\tensor \C(v_2)$$
Most of the theory developed here would look much nicer if it were.
Perhaps this is an indication that we are using slightly the wrong
defintion of $\C$.  However, if $v_1$ is of degree $1$ in $\V$ then we
can actually show the multiplicative property due to our assumption
that $\C(\Gamma)=0$ if $\Gamma$ is connected but not 1PI.

If $v_1$ has degree $1$ then any graph formed from $v_1\tensor v_2$ in
the process for computing $\C(v_1\tensor v_2)$ will either have a
single edge joining the vertex corresponding to $v_1$ to some vertex
from $v_2$ or it will have no such edge.  If the edge is there then
this graph contributes nothing to the action by the assumption on
$\C$.  Hence we see
$$\C(v_1\tensor v_2) = \C(v_1)\tensor \C(v_2)\quad\hbox{for $v_1$ of
degree $1$}$$

In fact, slightly more can be said.  It is easy to see that $\C(v_1) =
v_1$ for terms $v_1$ of degree $1$.  So, repeatedly using the above
multiplicative property we see that
$$\C(v_1\tensor\cdots\tensor v_n) = v_1\tensor\cdots\tensor v_n\quad
\hbox{for all $v_i$ of degree $1$}$$

Let $L$ be the interaction part of the Lagrangian (ie. the part with
the kinetic energy and mass terms removed).  It is then easy to see
that the terms occuring in the Green's function associated to some
choice of $v\in \lag(I)$ (these were basically the field that came
attached to $J$) were
$$\G(v) + \G(f_1^*(v\tensor L)) + {1\over 2!} \G(f_2^*(v\tensor L\tensor
L)) + \cdots$$
where $L$ is regarded as an element of $\lag(\{1\})$ and the maps $f_n$
are
$$f_n:I \longrightarrow I\sqcup\{1,2,\dots,n\}$$
We denote the above sum as
$$\G(v\tensor\exp(L))$$

Suppose we have a Lagrangian $L$ and a choice of renormalisation
prescription $\F$.  If we were to pick a second renormalisation
prescription $\F'$ then we know (from the previously proved
transitivity result) that there is a finite renormalisation $\C$ such
that $\F = \C[\F']$.  Recalling the commutative diagram:
\begin{diagram}[PostScript=dvips]
\lag(I) &\rTo& \feyn(I)\qquad & & \\
\dTo<{\C} & & \dTo<{\C} & \rdTo(2,1)^{\C[\F']}& \quad\dist(I)\\
\lag(I) &\rTo& \feyn(I)\qquad & \ruTo(2,1)_{\F'}& 
\end{diagram}
We see that the basic Green's function given by the two different
renormalisation prescriptions are related by
$$\G_{\F} = \G_{\F'} \circ \C$$
Hence the Green's functions from the Lagrangian are related by
$$\G_{\F}(v\tensor\exp(L)) = \G_{\F'}(\C(v\tensor\exp(L)))$$
Assume that $v$ is a tensor product of fields of degree $1$.  Then
this becomes
$$\G_{\F}(v\tensor\exp(L)) = \G_{\F'}(v\tensor \C(\exp(L)))$$
If we could find a new Lagrangian $L'$ such that
$$\exp(L') = \C(\exp(L))$$
Then we would have
$$\G_{\F}(v\tensor\exp(L)) = \G_{\F'}(v\tensor \exp(L'))$$
In other words, a change in renormalisation prescription can be
accompnied by a corresponding change in Lagrangian so that the Green's
function for the QFT are unchanged.  Remarkably this is possible.

Let $v_1\tensor\cdots\tensor v_n$ be a homogeneous term from $L^{\tensor
n}$.  Let $\gamma$ be a connected graph on $n$ vertices with the edges
at vertex $i$ being labeled by degree $1$ terms from $v_i$ and the
vertex being labeled with the remaining terms.  Now form the sum
$$v_1\tensor\cdots\tensor v_n \rightarrow \sum_\gamma
\C(\gamma)\gamma$$
where the action of $\C(\gamma)$ on a graph with labeled vertices has
already been explained.  We contract the resulting graph and multiply
all labels on the vertices.  This gives us an element of $\V$ becuase
$\gamma$ was connected.  The new Lagrangian is the sum of all such
terms over all possible $n$.

By taking the graph $\gamma$ to be the graph on one point we see that
$L'$ contains the original Lagrangian $L$.

Note that this definition of $L'$ is generally an infinite sum.  To
make this converge we regard the coupling constants (such as
$\lambda$) as being formal variables.  It is easy to see that there
will only be a finite number of terms of each degree in the coupling
constants and fields.  This, of course, now means that the new
Lagrangian technically isn't in the space of Lagrangians.  To get
around this we redefine the space of Lagrangians to be formal power
series in the coupling constants with coefficients in the former space
of Lagrangians.  Note that this isn't the same as $\CCC[[{\rm
coupling\ constants}]] \tensor_\CCC \V$.

\subsubsection{Integrating Distributions}

Finally, we ought to discuss how to integrate the resulting
distributions over a non--compact space.  If we have a distribution
$M_f$ coming from a function $f$ then its integral should be equal to
the integral of $f$ whenever it is defined.  In other words
$$\int M_f dx = \int f dx = \int f.1 dx = M_f(1)$$
Unfortunately, we can't in general apply a non--compactly supported
distribution to the constant function.  Look instead at the Fourier
transform.  This is sensible because we know
$$\hat{f}(0) = \int f dx$$
Now we can calculate
$$\hat{f}(0) = \int \delta(x) \hat{f}(x) dx =
\widehat{M_f}(\delta_x)$$
So, provided that $0$ is not in the singular support of $\hat{f}$ we
will be able to define the integral.  It is easy to see how this will
generalise to distributions.  So, what remains to be shown is if $0$
is not in the singular support of the distributions that occur.

TO BE CONTINUED...

\subsection{Finite Dimensional Orbits}

Given any Lagrangian $\L$ there are four obvious questions that we can
ask about it
\begin{enumerate}
\item What is the group of symmetries of $\L$?
\item Given these symmetries, is $\L$ the most general Lagrangian
fixed by them?
\item Can we find renormalisation prescriptions that are invariant
under some of these symmetries?
\item Is this space of renormalisation prescriptions acted on
transitively by finite renormalisations?
\end{enumerate}
In the above the ``group'' of symmetries should be allowed to include
things like derivations and supersymmetries (hence it technically
isn't a group).  Similarly, ``invariant'' and ``fixed'' should allow
surface terms and generalised eigenvectors.

If we are really lucky then the set of Lagrangians fixed by the
symmetries will be finite dimensional.  If this happens then there is
at least a hope of being able to test the theory by experiment (only a
finite number of experiments should be needed to determine the
constants whereas if there was an infinite dimensional space of
theories then one would never be able to determine the constants).

Normally it is difficult to arrange that the orbit in the space of
Lagrangian is finite dimensional.  One way to help make this more
likely is to assign a {\bf degree} to each Feynman diagram and use
this to place restrictions on the degree of the operators obtained
from these diagrams\footnote{Physicists call the degree a
dimension.}.  We assign the degree of $\d_\mu$ to be $1$ and the
degree of $x_\mu$ to be $-1$.  We then require that the Lagrangian
$\L$ have total degree $n$ in $n$--dimensional spacetime (this then
ensures that when we integrate $\L$ over spacetime the integral will
have degree $0$).

If we have a graph $\Gamma$ with no internal vertices then we assign
its degree as the sum of the degrees of the labels on its edges.  If
there are internal vertices then we subtract the dimension of
spacetime for every internal vertex (because they are integrated
over).

We would like to force the degree of $\F(\Gamma)$ to be
$-\deg\Gamma$.  Unfortunately this is not possible because massive
propagators are non-homogeneous in general.  So, instead we insist
that $\F(\Gamma)$ be a sum of terms of the form `a smooth function
multiplied by distributions of generalised homogeneous degree at least
$-\deg\Gamma$'.

We now have a space of renormalisation prescriptions.  We need to find
a space of finite renormalisation acting on these.  A little work
shows that the condition required is
$$\deg\C(\Gamma) \le \deg\Gamma - n({\rm vertices} - {\rm
components})$$
The reason for the strange looking extra factor is that this exactly
gives the degree of the distribution $\C(\Gamma)\delta({\rm diag})$.

As an example, consider the $\varphi^4$--theory in $4$--dimensional
spacetime.  The action of $\F$ on the Lagrangian will give terms of
the form $\C(\Gamma)\Gamma_{\rm vertex}$ where $\Gamma$ is some graph
with vertices and edges labeled by fields such that the fields at each
vertex combine to give $\varphi^4$; $\Gamma_{\rm vertex}$ is the
fields on the vertices which is acted on naturally by the differential
operator $\C(\Gamma)$.  The degree of this term is easy to estimate
(assuming $\Gamma$ is connected)
\begin{eqnarray*}
\deg(\C(\Gamma)\Gamma_{\rm vertex}) &=& \deg(\C(\Gamma)) + \sum
\deg\varphi^* \\
&\le& \deg\Gamma - 4(v-1) + \sum \deg\varphi^* \\
&=& 4v - 4(v-1) \\
&=& 4
\end{eqnarray*}
The equality in the third line follows because the fields at each
vertex total to $\varphi^4$ which has degree $4$.  Hence the action of
the renormalisation only generates terms of degree at most $4$.  Hence
there is a finite dimensional orbit.

If we do the exact same calculation but assume that the dimension of
spacetime is $n$ and the Lagrangian starts with terms of degree at
most $m$ then we find that $\F$ can generate terms with degree at most
$(m-n)v + n$.  To remain with a finite number of terms we clearly
require that $m\le n$.  Thus we should require that all terms in the
Lagrangian should have degree at most the dimension of spacetime.  In
terms of the coupling constants this translates to all coupling
constants having degree at least $0$.  This condition is called {\bf
Dyson's condition}.

It is easy to check that Dyson's condition holds for the QED
Lagrangian but doesn't hold for Lagrangians involving gravity.  So,
quantum gravity is hard to deal with in our current formulation (it is
a so called ``non-renormalisable'' theory).

Assume the dimension of space--time is $d$.  All our Lagrangians have
the term $\d^i\varphi \d_i\varphi$.  This term occurs with no coupling
constant and so it must have total degree $d$.  Hence
$$\deg(\varphi) = {d \over 2} - 1$$
If we are to have a term of the form $\lambda_k \varphi^k$ then we
know
$$\deg(\lambda_k) + k\({d \over 2} -1\) = d$$
hence
$$k\({d \over 2} - 1\) \le d$$
Solutions to this are easy to enumerate
$$\begin{array}{|r||c|l|}
\hline
d=2 & k \hbox{ is arbitrary} & \\
d=3 & k\le 6 & \hbox{this is $\varphi^6$--theory}\\
d=4 & k \le 4 & \hbox{this is $\varphi^4$--theory}\\
d=5 & k \le 3 & \hbox{the term $\varphi^3$ has non--peturbative
problems} \\
d=6 & k\le 3 & \hbox{as above}\\
d\ge 7 & k\le 2 & \hbox{this is free field theory}\\
\hline
\end{array}$$

\eject\section{Fermions}

In this section we study the mathematics behind the Dirac equation for
the electron.

We want to find a relativistic equation for the electron; we already
have the Klein--Gordon equation $\d^i\d_i \varphi = m^2 \varphi$.  We
want to write
$$(\d^i\d_i - m^2) = \(\sum A^i\d_i + m\) \(\sum A^i\d_i - m\)$$
If we expand this out we see that $2A^iA^j = g^{ij}$ and as our choice
of metric had $g^{ij}=0$ for $i\ne j$ this has no solutions in
dimensions greater that $1$.  However, if we suppose that $\varphi$ is
not a scalar valued function but takes values in some vector space $V$
then we are allowed to choose the $A^i$ as endomorphisms of $V$.  Thus
we get the equation
$$A^iA^j + A^jA^i = 2g^{ij}$$
Dirac found $4\times 4$ matrices which satisfied this equation.

Mathematically, this is a special case of the following problem.  Let
$E$ be a vector space over some field $k$ (of characteristic not $2$)
and let $q$ be a quadratic form on $E$.  Then, $q$ comes form a
symmetric bilinear form $(,)$
$$q(x) = (x,x) \qquad\hbox{ and }\qquad (x,y) = {1 \over 2}(q(x+y) -
q(x) - q(y))$$
We want to find matrices $A(x)$ for each $x\in E$ such that $A(x)^2 =
q(x)I$.\footnote{So, we are trying to find an associative algebra that
encodes the data of the quadratic form.  This is similar to the
definition of the universal enveloping algebra as being an associative
algebra encoding the data of a Lie bracket.}

\subsection{The Clifford Algebra}

Define the {\bf Clifford algebra} $\Clif(q)$ of a quadratic form to be the
algebra generated by $E$ with relations $x^2=q(x)$.  Then modules over
$\Clif(q)$ are matrices satisfying the relations we wanted.

Note that there is a natural $\ZZZ/2\ZZZ$ grading on $\Clif$ given by the
number of elements of $E$ in the product (this is well defined as the
only relation is $x^2=q(x)$).  Denote this decomposition as $\Clif = \Clif^0
\oplus \Clif^1$.

If $x_1,\dots,x_n$ forms an orthogonal basis for $E$ then the $2^n$
products $x_{i_1}\cdots x_{i_k}$ for $i_1 < \cdots < i_k$ span the
Clifford algebra.  So $\dim(\Clif) \le 2^n$.  It is easy to see that this
is an equality if $n=1$.

Note that if $E_1$ and $E_2$ are vector spaces with quadratic forms
$q_1$ and $q_2$ then
$$\Clif(E_1 \oplus E_2) \cong \Clif(E_1) \hat\tensor \Clif(E_2)$$
where $\hat\tensor$ is the graded tensor product
$$(a\hat\tensor c)(b \hat\tensor d) = (-1)^{\deg(c)\deg(b)} (ab
\hat\tensor cd)$$
By diagonalisation we can decompose $E$ as a direct sum of
$1$--dimensional spaces and hence we see that $\dim(\Clif) = 2^n$ and the
above products are a basis for $\Clif$.

There are $3$ natural automorphisms of $\Clif$
\begin{enumerate}\parskip 0pt
\item {\bf Transpose} which is an anti-automorphism $x \mapsto x^t$
induced by $x_1\cdots x_k \mapsto x_k\cdots x_1$
\item {\bf Negation} which is an automorphism $x \mapsto -x$.
Sometimes it is denoted by $\alpha(x)$.
\item {\bf Conjugation} which is an anti-automorphism given by $\bar x
= -x^t$.
\end{enumerate}

In order to study the representations of $\Clif$ it is natural to first
study the structure of the centre $Z(\Clif)$.  This is easy to work out
\begin{itemise}\parskip 0pt
\item If $n$ is even then $Z(\Clif)$ is one dimensional spanned by $1$
\item If $n$ is odd then $Z(\Clif)$ is two dimensional spanned by $1$ and
$x_1\cdots x_n$
\item If $n$ is even then $Z(\Clif^0)$ is two dimensional spanned by $1$
and $x_1\cdots x_n$
\item If $n$ is odd then $Z(\Clif^0)$ is one dimensional spanned by $1$
\end{itemise}

Suppose that the field $k=\CCC$.  Then we can find an orthonormal
basis $x_1,\dots , x_n$ for $E$.  Let $G$ be the subgroup of $C$
consisting of elements $\pm x_{i_1}\cdots x_{i_k}$.  $G$ has order
$2^{n+1}$.  Let $\varepsilon = -1$ which is an element of $G$.  Then
it is easy to see
$$\Clif(E) = \CCC[G] / (\varepsilon + 1)$$
Thus representations of $G$ with $\varepsilon$ acting as the scalar
$-1$ are the same as representations of $C$.  The number of
irreducible representations of a finite group $G$ is the number of
conjugacy classes.  It is easy to see that these are given by elements
of the centre $Z(G)$ and pairs of elements $\{x,-x\}$ for $x\not\in
Z(G)$.  Hence
\begin{itemise}\parskip 0pt
\item If $n$ is even there are $2^n + 1$ irreps
\item If $n$ is odd there are $2^n + 2$ irreps
\end{itemise}
For any Clifford algebra there are $2^n$ obvious $1$-dimensional
representations.  Therefore, if $n$ is even there is one more
irreducible representation and it must have dimension $2^{n/2}$ (from
the fact that the sum of squares of dimensions of irreps is the order
of the group).  If $n$ is odd then there are two more irreducible
representations to find.  These representations are swapped by the
involution $\alpha$ and hence each has dimension $2^{(n-1)/2}$.

If we restrict these irreducible representations to the subalgebra
$\Clif^0$ then the following occurs
\begin{itemise}\parskip 0pt
\item If $n$ is even then the unique irrep of dimension $2^{n/2}$
spits into a sum of two representations of dimension $2^{n/2 - 1}$
\item If $n$ is odd then the 2 irreps of dimension $2^{(n-1)/2}$
become isomorphic when restricted to $\Clif^0$.
\end{itemise}

The {\bf Clifford group} is defined to be
$$\Gamma = \{ x\in C : \alpha(x)yx^{-1} \in E\hbox{ for all $y\in E$}
\}$$
This is the group which when acting by twisted conjugation preserves
$E$.  It is easy to see that this includes all non-isotropic elements of
$E$ and the twisted conjugation acts by reflection.

The homomorphism from $\Gamma$ to $k^\times$ given by $N(x) = x\bar x$
is called the {\bf spinor norm}.  It is easy to see that $N(xy) =
N(x)N(y)$ for $x$,$y\in \Gamma$.  The value of the spinor norm on
elements of $E$ is preserved under twisted conjugation:
$$N(\alpha(x)yx^{-1}) = N(y)\qquad\hbox{ for all $x\in\Gamma$ and
$y\in E$}$$
Hence, the action of $\Gamma$ on the vector space $E$ is exactly the
orthogonal group.  We can define the action of the spinor norm on
element of the orthogonal group by choosing any lift of them to the
group $\Gamma$.  This is well defined in the group $k^\times /
(k^\times)^2$.  For example
\begin{itemise}\parskip 0pt
\item If $k=\CCC$ then $k^\times / (k^\times)^2 = 1$ and so all
reflections have spinor norm $1$.
\item If $k=\RRR$ and $q$ is positive definite then all reflections
have spinor norm $-1$.  Hence the spinor norm is the determinant.
\item If $q$ is negative definite then the spinor norm is always $1$.
\item If the form is indefinite ($\RRR^{m,n}$) then reflections in
positive norm vectors have spinor norm $-1$ and reflections in
negative norm vectors have spinor norm $1$.  So, spinor norm and the
determinant give two {\it different} homomorphisms from $\lieO(E)$ to
$\ZZZ/2\ZZZ$.  Thus the orthogonal group has at least $4$ components.
\item If there is one negative direction ($\RRR^{n-1,1}$) then the $4$
components are easy to see.  They are given by the determinant and
whether the two cones are swapped.
\item Over finite fields of characteristic not $2$, $k^\times /
(k^\times)^2 \cong \ZZZ/2\ZZZ$.
\item Over $\QQQ$, $k^\times / (k^\times)^2 \cong \ZZZ/2\ZZZ \times 
\ZZZ/2\ZZZ \times \cdots$
\end{itemise}

The {\bf Pin group} is defined to be
$${\rm Pin} = \{x\in \Gamma : N(x) =1 \}$$
This is a double cover of the orthogonal group.  The elements of this
which act with determinant $+1$ form the {\bf spin group}.  Note that
the spin group does not necessarily cover the special orthogonal group
because the spinor norm is always $1$.  We can alternatively describe
the spin groups as ${\rm Pin} \cap \Clif^0$.  Hence, representations of
the even Clifford algebra give rise (by restriction) to
representations of the spin group.

\subsection{Structure of Clifford Algebras}

All real Clifford algebras are sums of matrix groups over $\RRR$,
$\CCC$ and $\HHH$.  To completely describe the structure we will need
the following facts
\begin{itemise}\parskip 0pt
\item $\RRR \tensor \hbox{anything} = \hbox{anything}$
\item $\CCC \tensor \CCC = \CCC \oplus \CCC$
\item $\HHH \tensor \CCC = \mx_2(\CCC)$ by the Pauli matrices
\item $\HHH \tensor \HHH = \mx_4(\RRR)$ by the action $(x\tensor y)z =
xz\bar y$
\item $\mx_n(A) \tensor B = \mx_n(A\tensor B)$
\item $\mx_m(\mx_n(A)) = \mx_{mn}(A)$
\end{itemise}
We also need to know how to tensor the algebras $\RRR$, $\CCC$ and
$\HHH$:
$$\begin{array}{|c|ccc|}
\hline
\tensor&\RRR&\CCC&\HHH \\
\hline
\RRR&\RRR&\CCC&\HHH \\
\CCC&\CCC&\CCC\oplus \CCC&\mx_2\CCC \\
\HHH&\HHH&\mx_2\CCC&\mx_4\RRR\\
\hline\end{array}$$
The small Clifford algebras are easy to work out by hand:
\begin{itemise}\parskip 0pt
\item $\Clif(\RRR^{0,0}) = \RRR$
\item $\Clif(\RRR^{1,0}) = \RRR[x]/(x^2-1) = \RRR + \RRR$
\item $\Clif(\RRR^{0,1}) = \RRR[x]/(x^2+1) = \CCC$
\item $\Clif(\RRR^{2,0}) = \mx_2(\RRR)$
\item $\Clif(\RRR^{1,1}) = \mx_2(\RRR)$
\item $\Clif(\RRR^{0,2}) = \HHH$
\end{itemise}
There are recurrence relations which relate higher dimension Clifford
algebras to lower ones (be very careful with the indices)
\begin{itemise}\parskip 0pt
\item $\Clif(\RRR^{m+2,n}) = \mx_2(\RRR) \tensor \Clif(\RRR^{n,m})$
\item $\Clif(\RRR^{m+1,n+1}) = \mx_2(\RRR) \tensor \Clif(\RRR^{m,n})$
\item $\Clif(\RRR^{m,n+2}) = \HHH \tensor \Clif(\RRR^{n,m})$
\end{itemise}
The reason for the change in index order is because of the difference
the graded tensor product makes.  For an explicit decomposition we can
let $e_1$ and $e_2$ be basis elements for $\RRR^{2,0}$ (or the other
cases) and then pick basis elements $e_1e_2f_i$ for the $\RRR^{m,n}$
Clifford algebra (note that this makes the tensor product un-graded
and changes the norm).

Note that these imply
$$\Clif(\RRR^{m+8,n}) = \Clif(\RRR^{m,n+8}) = \mx_{16}(\Clif(\RRR^{m,n}))$$
In other words, the qualitative behaviour of the Clifford algebra
depends only on the signature modulo $8$.  By using the recurrence
relations we see that the Clifford algebra is a matrix algebra over
the following rings
$$\begin{array}{|r||cccccccc|}
\hline
\hbox{\bf signature} & 0&1&2&3&4&5&6&7 \\
\hline
\hbox{\bf ring} & \RRR&\RRR+\RRR&\RRR&\CCC&\HHH&\HHH+\HHH
&\HHH&\CCC\\
\hline
\end{array}$$

To work out the structure of the even Clifford algebra we only need to
notice that the full Clifford algebra can be generated by elements
$e_1, e_1e_2, e_1e_3, \dots, e_1e_n$ and the elements apart from $e_1$
generate $\Clif^0$.  However, they also form a Clifford algebra and so
\begin{itemise}\parskip 0pt
\item $\Clif^0(\RRR^{m+1,n}) = \Clif(\RRR^{n,m})$
\item $\Clif^0(\RRR^{m,n+1}) = \Clif(\RRR^{m,n})$
\end{itemise}
Hence the even Clifford algebra is a matrix algebra over the following
rings (except for the trivial even Clifford algebra)
$$\begin{array}{|r||ccccc|}
\hline
\hbox{\bf signature} & 0 &\pm 1&\pm 2&\pm 3&4 \\
\hline
\hbox{\bf ring} & \RRR+\RRR&\RRR&\CCC&\HHH&\HHH+\HHH \\
\hline
\end{array}$$

We can now explain some of the physics terminology about spinors.
\begin{itemise}\parskip 0pt
\item {\bf Dirac spinors} are elements of a complex representation of
a Clifford algebra.  Therefore they exist in all signatures $(m,n)$.
\item {\bf Weyl spinors} are elements of complex half spin
representations (i.e. the representation must split over $\Clif^0$).
These occur only in even signatures.
\item {\bf Majorana spinors} are elements of real spin
representations.  The above table shows these exist only in signatures
$0,1,2$ modulo $8$.
\item {\bf Majorana--Weyl spinors} are elements of real half spin
representations.  From the above table (and the condition that the
signature must be even) they occur only in signature $0$ modulo $8$.
\end{itemise}

\subsection{Gamma Matrices}

The gamma matrices are explicit matrices giving a representation of
the Clifford algebra $\Clif(\RRR^{m,n})$.  They are usually defined in
terms of the {\bf Pauli matrices}
$$\sigma^1 = \(\begin{array}{cc}0&1\\1&0\end{array}\)\qquad
\sigma^2 = \(\begin{array}{cc}0&-i\\i&0\end{array}\)\qquad
\sigma^3 = \(\begin{array}{cc}1&0\\0&-1\end{array}\)$$
The $4$--dimensional gamma matrices are then
$$\gamma^0 = \(\begin{array}{cc}I&0\\0&I\end{array}\)\quad
\gamma^1 = \(\begin{array}{cc}0&\sigma^1\\-\sigma^1&0\end{array}\)\quad
\gamma^2 = \(\begin{array}{cc}0&\sigma^2\\-\sigma^2&0\end{array}\)\quad
\gamma^3 = \(\begin{array}{cc}0&\sigma^3\\-\sigma^3&0\end{array}\)$$
The matrix $\gamma^5$ is defined to the the product of all the gamma
matices.  Note that the above matrices are a faithful representation
of the Clifford algebra $\Clif(\RRR^{1,3})$.

With this explicit realisation of the Clifford algebra we can
decompose $\Clif(\RRR^{1,3})$ under the action of $\lieO(\RRR^{1,3})$
$$\begin{array}{|c|c|c|}
\hline
\hbox{\bf space} & \hbox{\bf dim} & \hbox{\bf representation} \\
\hline\hline
1&1&\hbox{scalar} \\
\gamma & 4 & \hbox{vector} \\
\gamma\gamma & 6 & \hbox{tensor} \\
\gamma\gamma\gamma & 4 & \hbox{axial--vector = vector $\tensor$ det} \\
\gamma\gamma\gamma\gamma & 1 & \hbox{pseudo--scalar = scalar $\tensor$
det}\\
\hline
\end{array}
$$

Sometimes these representations are denoted by $S$, $V$, $T$, $PV$ and
$PT$.

Once we have chosen gamma matrices we can explain {\bf Feynman slash
notation}.  In physics books you often see operators (like $\d$) with
a slash through then (like $\FMslash{\d}$).  This is shorthand for
contracting against the gamma matrices.  So $\FMslash{\d} \psi$ is
shorthand for $\gamma^\mu \d_\mu \varphi$.

\subsection{The Dirac Equation}

Let us now explain the the fields and Lagrangian for the Dirac
equation.

There is a $8$--dimensional vector space of fields with basis given by
$$\psi_1, \psi_2, \psi_3, \psi_4, \psi_1^*,\psi_2^*, \psi_3^*,\psi_4^*
$$
These fields are interchanged in the obvious way under complex
conjugations.

For many of the formul\ae\ in physics books these fields are grouped
together into a vector with $4$ components.  So,
$$\psi = \(\begin{array}{c}\psi_1 \\ \psi_2 \\ \psi_3 \\
\psi_4\end{array}\) \quad \psi^* = \(\begin{array}{c}\psi_1^* \\
\psi_2^* \\ \psi_3^* \\ \psi_4^*\end{array}\)$$
The {\bf dagger operator} is then defined on these vectors as the
conjugate transpose
$$\psi^\dagger = \(\begin{array}{cccc} \psi_1^* & \psi_2^* & \psi_3^*
& \psi_4^*\end{array}\) \qquad {\psi^*}^\dagger = \(\begin{array}{cccc}
\psi_1 & \psi_2 & \psi_3 & \psi_4\end{array}\)$$
The {\bf bar operator} is defined by
$$\bar \psi := \psi^\dagger \gamma^0$$
The {\bf Dirac lagrangian} for the system is
$$\bar\psi (i\FMslash{\d} - m)\psi$$
The {\bf propagator} for this Lagrangian is easy to work out (because
it is something to do with the Klein--Gordon propagator).  It turns
out to be
$$(-i\FMslash{\d} + m)\Delta$$

\subsection{Parity, Charge and Time Symmetries}

There are three non-continuous symmetries that turn up all over
physics.  These are called $P$ (parity reversal -- space is reflected
in some mirror), $C$ (charge conjugation -- particles are replaced by
their anti-particles) and $T$ (time reverals -- the direction of time
is reversed).  Originally, it was believed that all three of these
were symmetries preserved by nature, however experiments have shown
this not to be the case.\footnote{The current belief is that the
product $PCT$ is a symmetry -- this can be proven provided you believe
the laws of QFT}  The formul\ae\ for $P$,$C$ and $T$ in physics books
are written down in terms of the gamma matrices.  At first sight they
look strange because they seem to involve the wrong gamma matrices
(why should time reversal involve $\gamma^1$ and $\gamma^3$ and not
$\gamma^0$?).  We will try to explain this in this section.  Firstly,
they formul\ae
$$\begin{array}{rclcrcl}
T\psi &=& -\gamma^1\gamma^3\psi &\qquad& T\bar\psi &=& \bar\psi
\gamma^1\gamma^3 \\
C\psi &=& -i\gamma^2\psi^* &\qquad& C\bar\psi &=& (-i\gamma^0 \gamma^2
\psi)^T \\
P\psi &=& \gamma^0\psi &\qquad& P\bar\psi &=& \bar\psi\gamma^0
\end{array}$$
These define the operators on a basis for the space of fields.  $P$
and $C$ are extended linearly and $T$ is extended anti-linearly to the
whole space of fields.

By simple computations one can show that $P$, $C$ and $T$ commute with
the conjugation $*$ and $\bar\psi\psi$ is invariant.  This suggests
that we should try to work in a space of fields $\Phi$ such that all
symmetries act projectively on this space.  As we are dealing with
spinors we need to find a central extension of the orthogonal group
acting on space--time.  This however can cause problems because for
non--connected groups the central extension is not always unique.  For
example, $\lieO(3,1)$ and $\lieO(1,3)$ are naturally isomorphic but
${\rm Pin}(3,1)$ and ${\rm Pin}(1,3)$ are not (look at the order of
elements).

We know that $\Clif(\RRR^{3,1}) \cong \mx_4\RRR$.  As the signature of
space--time is $2$ we know that Majorana spinors exist, so there is
some $2$--dimensional real vector space $S$ acted on by the Clifford
algebra.  Using the fact that matrix algebras over $\RRR$ have no
outer automorphisms we see that there is a unique (up to scalar
multiples) bilinear form on $S$ such that $(Ca,b) = (a,\bar Cb)$.

Does the Clifford algebra preserve this bilinear form?  Almost, a
simple calculation shows
$$(Ca,Cb) = (a,\bar CCb) = \bar CC(a,b)$$
So the form is multiplied by the spinor norm of $C$.

Now look at the representation $S\oplus S^*$ where the action of the
pin group is defined to be the usual one on $S$ and twisted by
$N(\cdot)$ on $S^*$.  Define the inner product by
$$(s_1\oplus s_2,s_3\oplus s_4) = (s_1,s_4) + (s_2,s_3)$$
It is then easy to check that this bilinear form is preserved under
the action of Pin.

Finally, let $\Phi = (S\oplus S^*)\tensor \CCC$.  This is acted on by
the Pin group in a way that preserves the bilinear form.  There is
also a natural conjugation action sending $S$ to $S^*$ and acting by
complex conjugation on $\CCC$.  We can define the charge conjugation
operator $C$ to be this conjugation $*$ extended linearly.

There is now a problem.  The orthogonal group does not commute with
the conjugation (an extra factor of $-1$ occurs for reflections of
spinor norm $-1$).  This can be fixed by letting elements of spinor
norm $-1$ act by multiplication by $i$ too.

We now have a representation of the spin group with a commuting action
of a symmetry called $C$.  The symmetries $P$ and $T$ are found inside
the spin group --- $T$ is a reflection perpendicular to the time axis
and $P$ is a reflection through the time axis.  Using these
definitions we will be able to write down explicit formulae for the
symmetries $C$, $P$ and $T$.  Why then do they seem to involve strange
combinations of the gamma matrices?

We want an inner product such that the adjoint of $x$ is $\bar x$.
This means we need to find a symmetric matrix with certain properties
and for the specific choice of gamma matrices made by physicists this
symmetric matrix happens to be equal to $\gamma^0$.

Why does charge conjugation act as $C\psi = -i\gamma^2 \psi^*$ rather
than the more obvious guess $C\psi = \psi^*$?  The answer is that
physicists choose a non-real basis for the vector space (because it
makes certain other calculation nicer) and hence the extra factor is
basically the matrix required to change the chosen basis to a real
one.

\subsection{Vector Currents}

If we pick for our space of fields $\Phi = (S\oplus S^*)\tensor \CCC$
then the space of possible terms in the Lagrangian is of the form
${\rm Sym}^\bullet (D\Phi)$.  An obvious choice for the Lagrangian of the
Dirac theory is
$$\L = \bar\psi i \FMslash{\d} \psi - m\bar\psi\psi$$
Unfortunately, this Lagrangian is not Hermitian.  There are two common
ways around this.  The first is to simply add the hermitian conjugate
to the Lagrangian (denoted in physics books by the symbols +h.c.).
The second way is to notice that it is hermitian up to surface terms
(total derivatives) and as the physics should not be affected by the
introduction of surface terms we can just work with the above
Lagrangian provided we remember that we may need to modify things by
surface terms.

As well as the Lagrangian being invariant under the obvious symmetries
(the rotation group, $C$, $P$ and $T$) there is also an slightly less
obvious global guage symmetry.  This is given by
$$\psi \mapsto e^{i\theta} \psi \qquad \bar\psi \mapsto e^{-i\theta}
\bar\psi$$
It is easy to check that this is a symmetry.  In its infinitessimal
form it is
$$\delta\psi = i\theta\psi \qquad \bar\psi = -i\theta\bar\psi$$
The conserved current is easy to work out with methods from the first
half of the course and is
$$j^\mu = \bar\psi\gamma^\mu \psi$$
It is easy to check that this current transforms as a vector and hence
it is called the {\bf vector current}.  If there is no mass term
(i.e. $m=0$) then there is another symmetry given by
$$\psi \mapsto e^{i\theta\gamma^5} \psi \qquad \bar\psi \mapsto
e^{-i\theta\gamma^5}\bar\psi$$
The conserved current in this case is
$${j^5}^\mu = \bar\psi\gamma^\mu\gamma^5 \psi$$
This time it transforms as an axial vector and hence is called the
{\bf axial vector current}\footnote{This is an example of a current
that is preserved in the classical case but can end up being
spontaneously broken in the quantum theory as renormalisation might
introduce a mass term for the electron.  This is called the axial
vector anomaly}.  

\end{fmffile}


\end{document}